\documentclass{aa}

\usepackage{amssymb}
\usepackage{amsmath}
\usepackage{graphicx}
\usepackage{url,natbib,twoopt}
\usepackage[varg]{txfonts}
\usepackage{subcaption}
\usepackage{todo}
\usepackage{hyperref}

\usepackage{ifxetex,ifluatex}

\hypersetup{
    colorlinks=true,
    linkcolor=blue,
    urlcolor=blue,
    citecolor=blue,
}

\bibpunct{(}{)}{;}{a}{}{,}
\graphicspath{{./figs/}}

\newcommand{\kev}{\ensuremath{\,\text{keV}}\xspace}
\newcommand{\cyc}{\ensuremath{\text{cyc}}}
\newcommand{\V}{\ensuremath{\text{V}}}
\newcommand{\gauss}{\ensuremath{\,\text{G}}\xspace}

\newcommand{\ns}{\mathrm{NS}}
\newcommand{\el}{\mathrm{e}}
\newcommand{\T}{\mathrm{T}}

\newcommand{\msun}{\,M_{\odot}}
\newcommand{\diff}{\ensuremath{\mathrm{d}}}

\begin{document}

\title{Fingerprints of thermal Comptonization in accreting neutron stars}
\subtitle{Plasma--vacuum interplay in cyclotron lines and polarisation}

\titlerunning{Thermal resonant Comptonization}  

\author{
        E.~Sokolova-Lapa\inst{1}\corrauth{ekaterina.sokolova-lapa@fau.de}
        \and
        D.~K.~Maniadakis\inst{2,3}\email{dimitrios.maniadakis@inaf.it}
        \and
        J.~J.~R.~Stierhof\inst{1}\email{jakob.stierhof@fau.de}
        \and
        E.~Ambrosi\inst{2}\email{elena.ambrosi@inaf.it}
        \and
        A.~D'A\`i \inst{2}\email{antonino.dai@inaf.it}
        \and
        C.~Ferrigno\inst{4,5}\email{carlo.ferrigno@unige.ch}
        \and
        N.~Schettino\inst{1}\email{nureldinschettino@gmail.com}
        \and
        M.~Middleton\inst{6}\email{M.J.Middleton@soton.ac.uk}
        \and
        A.~G{\'u}rpide\inst{6, 7}\email{a.gurpidelasheras@uva.nl}
        \and
        V.~Grinberg\inst{8}\email{grinberg.astro@gmail.com}
        \and
        G.~Lipunova\inst{1, 9}\email{galina.lipunova@fau.de}
        \and
        I.~El Mellah\inst{10}\email{ileyk.elmellah@upc.edu}
        \and
        P.~Kretschmar\inst{11}\email{peter.kretschmar@esa.int}
        \and
        J.~Wilms\inst{1}\email{joern.wilms@fau.de}
}

\institute{
    Dr.~Karl Remeis-Observatory and Erlangen Centre for Astroparticle Physics, Friedrich-Alexander Universit\"at Erlangen-N\"urnberg, Sternwartstr.~7, 96049 Bamberg, Germany
    \and
    INAF - IASF Palermo, via Ugo La Malfa 153, 90146 Palermo, Italy
    \and
    Dipartimento di Fisica e Chimica Emilio Segr\`e, Universit\`a degli Studi di Palermo, Via Archirafi 36, 90123 Palermo, Italy
    \and
    Department of Astronomy, University of Geneva, Chemin d'\'Ecogia 16, CH-1290 Versoix, Switzerland
    \and
    INAF, Osservatorio Astronomico di Brera, Via E. Bianchi 46, I-23807 Merate, Italy
    \and
    School of Physics and Astronomy, University of Southampton, University Road, Southampton SO17 1BJ, UK
    \and
    Anton Pannekoek Institute for Astronomy, University of Amsterdam, Science Park 904, 1098 XH Amsterdam, The Netherlands
    \and
    European Space Agency (ESA), European Space Research and Technology Centre (ESTEC), Keplerlaan 1, 2201 AZ Noordwijk, The Netherlands
    \and
    Max-Planck-Institut f\"ur Radioastronomie, Auf dem H\"ugel 69, 53121
    Bonn, Germany
    \and
    Departament de Física, EEBE, Universitat Politècnica de Catalunya, c/ Eduard Maristany 16, 08019, Barcelona, Spain
    \and
    European Space Agency (ESA), European Space Astronomy Centre (ESAC), Camino Bajo del Castillo s/n, 28692 Villanueva de
la Ca{\~n}ada, Madrid, Spain
}

\abstract
  {
  X-ray emission from accreting, strongly magnetised
  neutron stars and its pulse-phase variability probe
  their magnetic-field geometry, spin orientation, and
  emission processes. Although the radiation likely
  originates in the lower part of the accretion channel,
  whether it emerges predominantly from a hot spot or an
  accretion column, and whether its properties are shaped
  primarily by bulk or thermal Comptonization, remain
  debated. We aim to disentangle intrinsic emission
  properties from geometrical visibility effects and
  identify observables characteristic of thermal
  Comptonization in hot-spot and column emission regions.
  We therefore derived their energy-dependent beam patterns
  and observable signatures without assigning the
  model to a particular luminosity regime, focusing on
  cyclotron-resonance and polarisation effects as
  tracers of emission anisotropy. To do so, we computed
  angle-dependent polarised broadband spectra, including
  the fundamental cyclotron line, for a homogeneous,
  self-emitting, strongly magnetised Comptonizing
  plasma and explored a broad range of physical
  parameters. Accounting for relativistic light
  bending and projection effects, we obtained
  phase-dependent fluxes for different geometries
  and, for the hot-spot models, observed linear
  polarisation. The beam patterns show a well-defined
  evolution with energy, driving characteristic changes
  in the pulse profiles. Near the cyclotron resonance,
  plasma--vacuum interplay produces a narrow central
  beam and extended side petals. Their changing visibility
  creates geometry-dependent dips, bumps, and M- and
  W-shaped structures in hot-spot pulsed-fraction
  spectra, but only dips and bumps for columns. We
  show that thermally Comptonized cyclotron lines do
  not reliably trace the plasma temperature and
  that plasma-induced ellipticity and energy-band
  averaging generally reduce the observed soft-X-ray
  linear polarisation to $0$--$30\%$ for typical
  parameters. Under conditions characteristic of
  X-ray pulsar accretion channels, thermal
  Comptonization leaves robust, energy-dependent
  anisotropic signatures: energy-resolved pulse
  profiles, pulsed fraction spectra, and
  polarisation provide complementary diagnostics
  of neutron star geometry, emission-region shape,
  and the spectral-formation mechanism.
  }

\keywords{X-rays: binaries – stars: neutron – methods: numerical
– Radiative transfer - Magnetic fields - Polarization}

\date{Received date / Accepted date }

\maketitle
\nolinenumbers

\section{Introduction}
\label{sec:intro}

The emission of accreting neutron stars in high-mass X-ray
binaries, also known as accreting X-ray pulsars, is powered
by accretion from their massive companions.
The strong magnetic field, ${\sim}10^{12}\gauss$, disrupts
the inflowing matter hundreds of gravitational radii from
the neutron star and channels it towards the magnetic poles,
confining accretion to only a small fraction of the stellar
surface. Gravitational energy released near the poles
powers X-ray beams that sweep across our line of sight (LOS) as
the star rotates, producing periodical flux variation on
typical timescales of a few to several thousand seconds.

The primary emission is expected to arise from dense,
decelerated plasma at the neutron star surface, where the
accreted matter merges into and heats up the upper
atmospheric layers, forming a ``hot spot''. The emitting
region may nevertheless extend vertically if supported by
a balance of magnetic, radiation, and gas pressure. Compton
scattering\footnote{Here and throughout, we use ``Compton
scattering'' broadly in the sense of Comptonization,
encompassing both direct and inverse scattering depending
on the electron and photon energy distributions.} energises
the radiation field and shapes the emergent spectrum.
It can proceed through two distinct channels: bulk and
thermal Comptonization, since scattering can occur on
fast-moving electrons in the free-falling bulk flow or on
hot electrons in the decelerated region
\citep[e.g.,][]{becker2007, farinelli2016}.
Their relative importance depends on the flow-deceleration
mechanism, anisotropic beaming of the radiation, and the
physical conditions within the accretion channel.

At present, there is no consensus on the detailed physics
of flow deceleration and emission across accretion regimes,
commonly characterised theoretically by the mass-accretion
rate, $\dot{M}$, and observationally by the X-ray luminosity.
Competing models differ in their assumptions about shocks,
instabilities, radiative properties, and the relative
efficiency of the processes governing the strongly magnetised
flow. A broad distinction can be made between low- and
intermediate-luminosity models and the models describing
the high-luminosity regime. For the former, proposed
scenarios include direct, thermally Comptonized emission
from an atmosphere, where the flow is stopped by Coulomb
or nuclear collisions
\citep{harding1984, miller1989, mushtukov2021, sokolova-lapa2021};
reprocessing of this radiation in the bulk flow
\citep{mushtukov2015b, fotiadis2026, markozov2026b}; and
an emitting column below a collisionless shock in the channel
\citep{langer1982, bykov2004, becker2022}.
High-luminosity models generally invoke a radiation-dominated
shock formed by a growing number of photon--electron
interactions that decelerate the bulk flow and energise
the radiation field \citep{davidson1973, basko1976}. These
models consider direct emission from the accretion column
\citep{rebetzky1989, gornostaev2025, falkner2026a, falkner2026b, gibson2026}, sometimes accompanied by atmospheric reprocessing
of downward-beamed column radiation, commonly called
``reflection'' \citep{kaminker1976, lyubarskii1988b, kraus1989,  poutanen2013, postnov2015}.

\begin{figure*}
  \centering
  \includegraphics[width=0.87\textwidth]{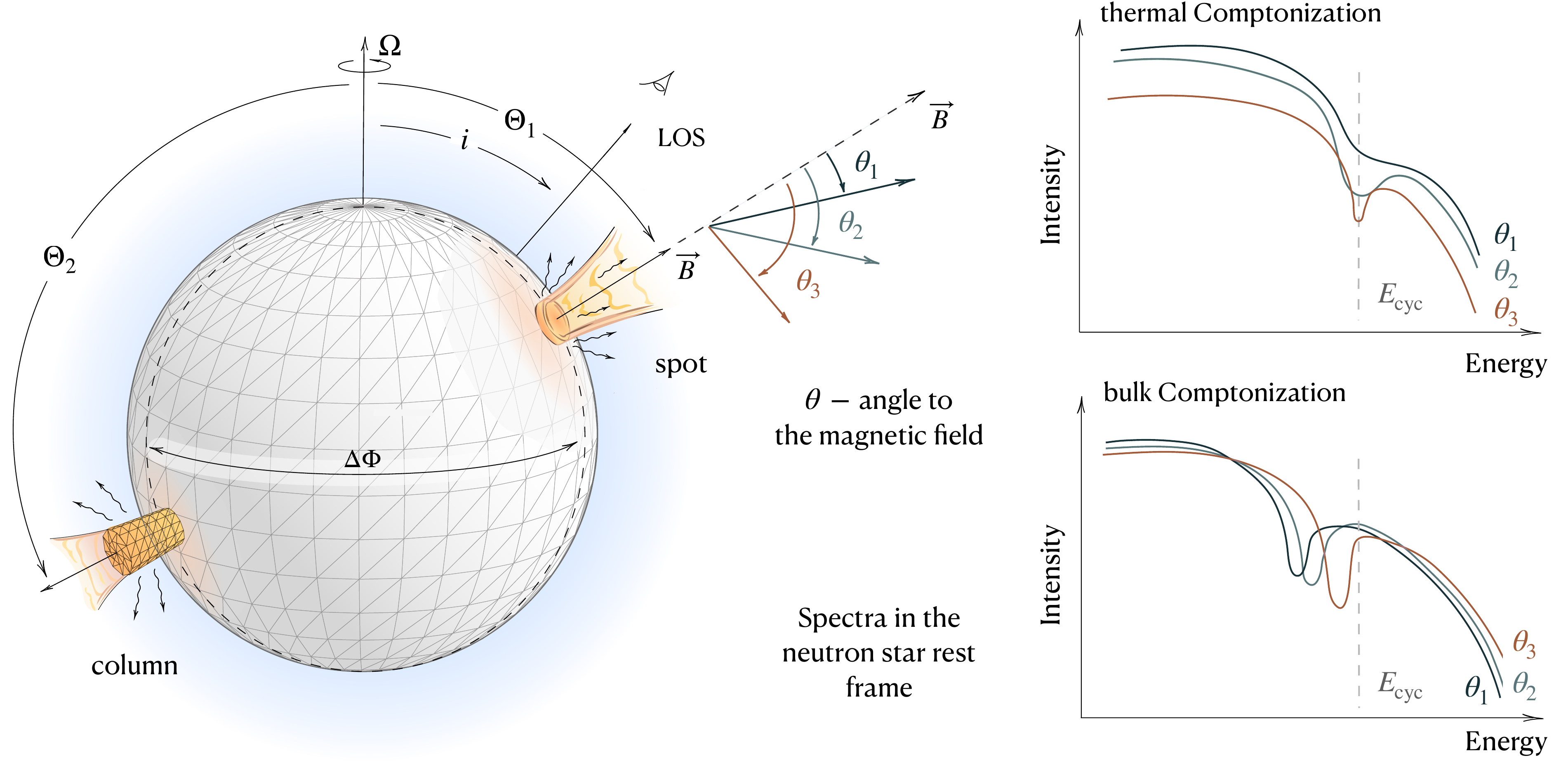}
  \caption{Left: Neutron star geometry, defined by the
  magnetic-pole angles, $\Theta_1$ and $\Theta_2$, and
  the LOS inclination, $i$, to the spin axis.
  Right: Spectra at different angles
  to the magnetic field, illustrating different
  beaming for thermal and bulk Comptonization.}
\label{fig:ns_tb}
\end{figure*}

The emission-region shape is often described through
the directionality of its beam pattern: a fan beam
produced by the side walls of a column or a pencil
beam from a hot spot
\citep{becker2012, mushtukov_tsygankov2023}. Although
fan and pencil beams are conventionally associated
with high- and low-luminosity regimes, respectively
\citep[e.g.,][]{markozov2024}, the relation
between beam pattern and luminosity is
more ambiguous. Thus, at low $\dot{M}$,
collisionless shocks may produce columns
\citep{vybornov2017, becker2022}. At
high luminosities, the illuminated atmosphere may
outshine the column, producing a pencil-like beam component.
Similar beaming may also arise from magnetohydrodynamic
(MHD) instabilities within the column, such as photon bubbles
propagating upstream through the accretion flow
\citep{klein1996, hsu1997, sheng2023}. MHD instabilities
or local thermal pressure exceeding magnetic confinement
may also cause matter to leak from the column and spread
across the surface \citep{mukherjee2013,abolmasov2023},
potentially favouring pencil-beam emission. Atmospheric
reprocessing, instability-driven
escape of radiation, and matter spreading may therefore
enhance spot-like emission relative to direct column
one, also diluting the observational signatures of
bulk Comptonization.
At a basic level, the debate centres on whether, for a
given source state, the
observed flux is dominant by a column-like (fan) or spot-like
(pencil) emission, and whether its spectral
properties are shaped primarily by bulk or thermal
Comptonization, that is, in the dynamic or steady-state
region.

Beyond controlling the accretion geometry, the
magnetic field also 
modifies plasma response, making photon--electron
interactions intrinsically anisotropic and
polarisation dependent. In such a medium, high-energy
radiation propagates in two normal polarisation modes
characterised by different refractive indices
\citep{gnedin1974}. Moreover, at magnetic field
strengths $B \gtrsim 10^{11}\,\gauss$, the transverse
motion of electrons is quantised into Landau levels,
whose characteristic separation, approximately given
by the cyclotron energy,
$E_\cyc \simeq 12\,\mathrm{keV} \times B_{12}$, with
$B_{12} = B / 10^{12}\,\mathrm{G}$, falls within the
X-ray energy range. Landau
quantisation fundamentally modifies radiative processes
in this regime by introducing further anisotropy and
strong resonances at the
cyclotron energy \citep{canuto1970, basko1975}. Its
most prominent observational manifestation is the
presence of broad absorption-like features in X-ray
spectra, known as cyclotron lines or cyclotron resonance
scattering features, whose energy allows for a direct
estimate of the magnetic field
\citep[see][and references therein]{staubert2019}.

The diagnostic potential of the cyclotron lines
is not limited to magnetic field estimates. Their
angular dependence can also probe the geometric
properties of the emission region
\citep{meszaros1988b, bulik1995} and the contributions
of different processes. In thermal- and bulk-scattering
scenarios line profiles vary strongly with viewing
angle, but their angular variations are different.
When the lines form in the decelerated plasma shaped
by thermal Comptonization, their centroid energy
remains nearly unchanged, but due to Doppler broadening
the width dramatically increases when the LOS approaches
the magnetic field direction
\citep[e.g.,][]{meszaros1985a,araya1999,schwarm2017a}.
When the lines are instead reprocessed in the bulk
flow, their centers appear shifted across a broad
energy range between different viewing directions,
while their widths may remain comparatively similar
\citep{rebetzky1989, fotiadis2026, falkner2026b}.
Figure~\ref{fig:ns_tb} (right) illustrates the angular
dependence in both cases.

The use of cyclotron lines as diagnostics for
emission-formation scenarios requires disentangling
the intrinsic variability of the line and continuum
from the visibility effects that include
gravitational light bending, contribution of several
emission regions, and the geometry: the location of
the regions on the surface, commonly
associated with the magnetic axis orientation, and
the inclination of the spin axis to the LOS (see
Fig.~\ref{fig:ns_tb}).

The strong magnetic fields modify not only
the plasma, but also the quantum-electrodynamic (QED)
vacuum, giving it an anisotropic dielectric response
and causing vacuum birefringence \citep{heisenberg1936}.
The plasma and vacuum contributions vary with photon
energy, plasma density, and magnetic-field strength.
Plasma birefringence is
defined by electrons and tends to make the normal
modes circularly polarised, while weakening the
cyclotron resonance for the mode whose electric
vector rotates opposite to the electron gyration.
Vacuum birefringence, by contrast, produces linear
modes that couple equally strongly to the resonances. 
Because these contributions produce different
energy-dependent anisotropies in the dielectric
response, both must be treated consistently to
predict the spectra, polarisation, and pulse-phase
variability 
\citep[e.g.,][]{sokolova-lapa2023phd, markozov2026b}.

The aim of this work is to identify directly observable
signatures of thermal Comptonization in the continuum,
cyclotron lines, polarisation signal, and their
variability with the rotational phase. We combined
polarised radiative transfer calculations,
including a detailed treatment of continuum and
resonant Comptonization in spot- and column-like
emission regions \citep{sokolova-lapa2023}, with a
three-dimensional model of a neutron star that accounts
for relativistic light bending, visibility, and projection
\citep{falkner2026a}.
We deliberately do not associate emission region shapes
with a luminosity regime, only specifying
a range of studied local plasma conditions.

Several observational results motivate our choice of
observables. In many sources, pulse profiles are
more complex at soft than at hard energies: a trend
that lacks a general physical explanation
\citep{alonso-hernandez2022,pottschmidt2026}.
Pronounced pulse-profile variations also occur near
cyclotron lines, where the pulsed fraction exhibits
substantial but diverse trends \citep{ferrigno2023, dai2025, maniadakis2025}.
Soft-X-ray polarimetry with the Imaging X-ray Polarimeter
Explorer \citep[IXPE,][]{weisskopf2022} further reveals
generally low, energy-dependent polarisation degree (PD) and,
in some sources, complex change of polarisation behaviour
between neighboring energy bands
\citep[][and references therein]{poutanen2024, forsblom2025, loktev2025, zhao2026}. Modelling remains largely
limited to semi-phenomenological descriptions of the
phase-dependent polarisation angle (PA), often broadly
consistent with the sky projection of the magnetic
axis. The PD remains poorly understood,
with low values sometimes attributed to the
temperature structure of the emitting region at intermediate luminosities
\citep{doroshenko2022, tsygankov2022}. We treat these
signatures not as independent phenomena, but as possible
manifestations of Comptonization in dense, hot, strongly
magnetised plasma near the neutron star surface, where
the competing plasma and QED-vacuum contributions jointly
shape the emergent radiation field. Section~\ref{sec:met}
introduces our modelling setup, Sects.~\ref{sec:res:sigelt}
and \ref{sec:res:lbsigelt} present the properties of the
internal emission and resulting observables for different
emission regions and geometries, and Sect.~\ref{sec:res:pol} describes the resulting soft X-ray polarisation. 
Section~\ref{sec:ende} summarises our findings and provides
an outlook.

\section{Problem formulation and methods}
\label{sec:met}

We modelled a strongly magnetised, slowly rotating
neutron star with two accreting poles, emitting from
surface hot spots or accretion columns. We first
computed internal emission for two medium orientations
(``slab'' and ``cylinder'') over a range of physical conditions (Sect.~\ref{sec:sigelt_mod}).
The resulting models were used for phase-dependent
relativistic projection calculations
(Sect.~\ref{sec:met:lbsigelt}) for hot spots and
column sidewalls, respectively. We generally assumed
antipodal dipolar poles, with mild asymmetries
considered in selected cases.

\subsection{Internal emission model}
\label{sec:sigelt_mod}

\subsubsection{General setup}
We modelled radiation emerging from a static,
strongly magnetised hydrogen plasma by solving
radiative transfer in the emission region.
The magnetic field is normal to the slab and
parallel to the cylindrical sidewall and axis.
We treated it as uniform because its dipolar variation
over the adopted vertical extent is
${\sim}7\%$ for the most extended model, remaining
below ${\sim}4\%$ in all other models and typically
below ${\sim}2\%$ (assuming the neutron star
radius of $R_\ns=12\,\mathrm{km}$).
The electron temperature and number density are
also constant, representing characteristic averages.
In reality, the plasma in accretion regions, subject
to a strong gravitational field and gradual
deceleration, is strongly inhomogeneous
\citep[see, e.g.,][]{zhang2022,gornostaev2025,sokolova-lapa2021},
with the temperature and density profiles depending
on the deceleration mechanism, mass-accretion rate,
and assumed efficiencies of relevant processes
\citep[see][]{becker2012,sheng2023,mushtukov_tsygankov2023}.
To remain agnostic about the accretion rate, we
therefore avoided prescribing a particular
structure for the emitting region.

\subsubsection{Radiative processes}

To describe the propagation of polarised radiation in
a strongly magnetised plasma, we adopted the formalism
of two normal modes \citep{gnedin1974, pavlov1979}.
Following \citet[][MN85a hereafter]{meszaros1985a},
we used photon polarisation vectors that include the
effects of a hot, non-relativistic plasma and the QED
vacuum. We also imposed a continuous behaviour
of the refractive indices for each mode across the
full energy range, and refer to them as ``mode 1''
and ``mode 2'' \citep[for a detailed discussion on the
mode choice and terminology see][and references
therein]{sokolova-lapa2023}.

\begin{figure}
  \includegraphics[width=\columnwidth]{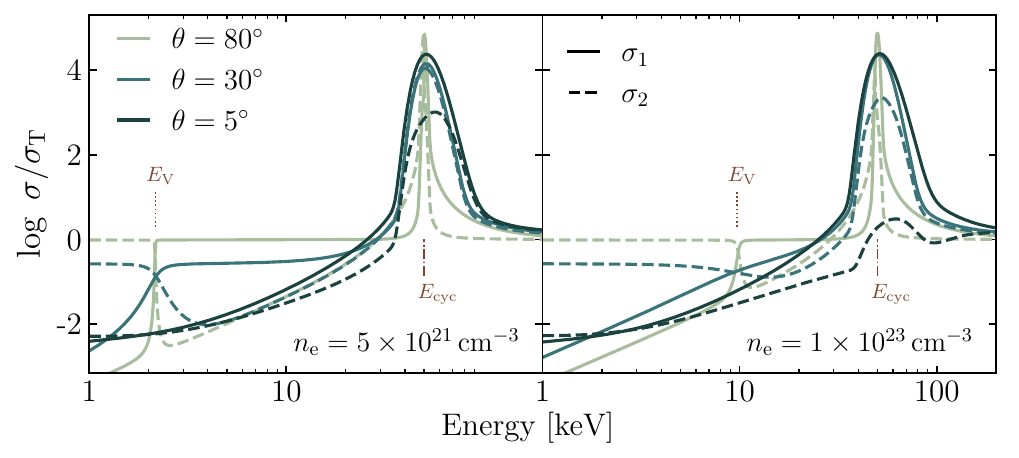}
  \caption{Compton scattering cross sections in a strong magnetic
           field for the two modes for $kT_\mathrm{e}=5\,\kev$ and
           $E_\cyc=50\,\mathrm{keV}$.
           }
\label{fig:crs}
\end{figure}

We assumed that the dominant emission processes are
bremsstrahlung from electron–proton collisions and
cyclotron emission due to the radiative decay of
electrons excited to higher Landau levels by these
collisions \citep[][]{nagel1980, arons1987}. The cyclotron
resonances are incorporated into the free–free
emissivity and absorption coefficients\footnote{
Under these conditions, the cyclotron emissivity
derived via Kirchhoff’s law substantially
overestimates the thermal photon production rate,
as collisions cannot maintain local
thermodynamic equilibrium
\citep[][and references therein]{nagel1983}.}
(\citealt{nagel1980, nagel1981b}, \citealt{nagel1983},
MN85a). The emission is Comptonized by anisotropic,
polarisation-dependent scattering on electrons
moving thermally along the magnetic field
lines. A characteristic feature of the corresponding
cross sections is a strong dependence
of the cyclotron-resonance profile on the angle
to the magnetic field, $\theta$ (e.g., MN85a,
\citealt{mushtukov2016}, \citealt{schwarm2017a}).
For all processes considered, we adopted the expressions
from MN85a, modifying only the detailed balance condition
to approximate stimulated scattering
\citep{meszaros1989, alexander1989}. 
Figure~\ref{fig:crs} shows an example of
scattering cross sections used here, with the
vacuum-resonance feature appearing at
\begin{equation}
\label{eq:ev}
  E_\V \approx 9.8 \kev \left( \frac{n_\el}{10^{23}\,\mathrm{cm}^{-3}} \right)^{1/2} \left( \frac{E_\cyc}{50\kev}\right)^{-1},
\end{equation}
where $n_\el$ is the electron number density in the
plasma \citep[see more details in][]{ventura1979, ho2003, sokolova-lapa2023}.

\subsubsection{Transfer and table model}

We solved the radiative transfer equation using the
\texttt{FINRAD} code
\citep{sokolova-lapa2021, sokolova-lapa2023phd}.
The code employs the Feautrier method and accounts
for photon redistribution in the continuum and at
the cyclotron resonance, solving for the specific
intensities of the two polarisation modes, $I_{1,2}(\mu,E)$,
where $\mu=\cos\theta$. The slab and cylinder emission
regions are represented by plane-parallel atmospheres
with normals $\vec{n}\parallel\vec{B}$ and $\vec{n}\perp\vec{B}$,
respectively.
The emergent intensities are obtained for $\mu\in[0,1]$
and extended to all outward directions by reflection about
$\vec{n}$ and azimuthal symmetry. \citep[see MN85a,][]{meszaros1992}.
The principal physical parameters are the
cyclotron energy, $E_\cyc$, the electron (longitudinal)
temperature, $kT_\el$, the electron number density,
$n_\el$, and the total Thomson optical depth, $\tau_\T$.
We parameterised the magnetic-field strength $B$ through
$E_\cyc$, which directly specifies the resonance
position and sets the energy scale.
The medium is self-emitting, with boundary conditions
of no incoming radiation at the top, $I(\mu<0, \tau=0)=0$,
and the diffusion limit at $\tau_\T$. We
computed a grid of models over the parameter ranges
specified in Table~\ref{tab:par} for both slab-like and
cylinder-like media. The resulting local, bin-integrated
photon specific intensities were combined into FITS tables
following the OGIP \texttt{XSPEC} table-model format
\citep{arnaud1996}, hereafter the \texttt{sigel-T}
local-emission tables (see Appendix~\ref{sec:data} for
model availability).
\begin{table}
\centering
\caption{Parameter ranges, units, and numbers of grid points,
$N_{\rm grid}$, used for the \texttt{sigel-T} model.}
\label{tab:par}
\begin{tabular}{lccc}
\hline
\hline
Parameter & Range & Unit & $N_{\rm grid}$\\
\hline
Cyclotron energy $E_\cyc$ & 20--100 & keV & 9\\
Temperature $kT_\el$ & 3--8 & keV & 6\\
Number density $n_\el$ & 1--50 & $10^{22}\,\mathrm{cm}^{-3}$ & 4\\
Optical depth $\tau_\T$ & 5--100 & & 5\\
\hline
\end{tabular}
\end{table}

\subsection{Observer's view}
\label{sec:met:lbsigelt}

\subsubsection{Geometry}
\label{sec:met:lbsigelt:geom}

We constructed a three-dimensional model of a neutron star
with two identical emitting regions using \texttt{lbscripts}
\citep{falkner2026a,falkner2026b}.
At each rotational phase, the code determines
the visible surface by tracing photon trajectories in the
Schwarzschild metric and computes the observed flux,
$F_E(\phi)$, including gravitational effects.
Each surface element, $\mathrm{d}S$, is assigned the same
local total intensity computed with \texttt{sigel-T},
$I(\mu,E)=I_1(\mu,E)+I_2(\mu,E)$. Slab and cylindrical
emission models are used for hot spots and column
sidewalls, respectively. The polar cap radius, $r_0$,
sets the transverse extent of the emitting region, while
the column height, $h$, additionally determines its
vertical extent.

The locations of the emission regions are specified by
their colatitudes relative to the spin axis, $\Theta_{1,2}$
and their azimuthal separation, $\Delta \Phi$. Together
with the observer inclination, $i$, these angles define
the system geometry (see Fig.~\ref{fig:ns_tb}). The
region with the smaller closest-approach angle to the
observer is the primary pole and the other is the
secondary pole. We generally considered symmetric
configurations with antipodal poles, $\Delta \Phi = 180^\circ$
and $\Theta_2 = 180^\circ - \Theta_1$. The latter
means that $\Theta_1$ gives
the magnetic inclination and the geometry is fully
described by the pair of angles $(i, \Theta_1)$. The
phase-dependent spectral flux is invariant to their
interchange, $F(E, \phi; i, \Theta_1) = F(E, \phi; \Theta_1, i)$.
In the following, we therefore denote these angles
by $(i_1, i_2)$ to define the geometry, except for
the PA angle modelling in Sect.~\ref{sec:res:pol}.
We also explored a few mildly asymmetric configurations
with $\Delta\Phi$ deviating from $180^\circ$ by up to ${\approx}20^\circ$.

\subsubsection{Compactness and observed emission}
\label{sec:met:lbsigelt:emang}

The strong anisotropy of the internal radiation makes the range of
emission angles, $\theta$, visible during the neutron star
rotation crucial to the observables. In principle, the spatial
extent of the regions allows radiation emitted at a range
of angles to contribute at each rotational phase.
In this work, we considered compact regions with opening
angles $\rho_0 \lesssim 5^\circ$, where
$\rho_0[\mathrm{rad}] \approx r_0 / R_\ns$ in the
small-angle approximation. Given the uncertain relation between the accretion
regime and emission region (see Sect.~\ref{sec:intro}), we
prescribed $\rho_0$ independently, not relating it to
$\dot{M}$ through the magnetospheric radius. This assumes
only that emission is dominated by compact regions near the
magnetic poles, which limits the range of angles simultaneously
contributing to $F_E(\phi)$.

For a given emission region configuration, the phase
dependence of the emission angle is determined by the
neutron star geometry and compactness. We defined the
compactness as $\mathcal{C} = R_\mathrm{s} / R_\ns$,
with the Schwarzschild radius $R_\mathrm{s}$.
The influence of high compactness is particularly
pronounced at large emission angles due to strong
light bending, which affects the visibility of the
secondary pole and its contribution to the observed flux.

We then computed $F_E(\phi)$ for selected combinations
of the neutron star compactnesses,
$\mathcal{C}=0.25$--$0.49$, polar cap radii,
$r_0=140$--$500\,\mathrm{m}$,
and column heights, $h=10$--$300\,\mathrm{m}$, and
the parameters of the internal emission listed in
Table~\ref{tab:par}. For $R_\ns=12\,\mathrm{km}$, the
$\mathcal{C}$ range corresponds to the neutron star
masses of ${\simeq}1$--$2\,\msun$. The adopted $r_0$
and $h$ describe compact near-surface regions, motivated
by two-dimensional simulations in which the quasi-steady,
directly emitting part of the column not embedded in the
surrounding inflow is typically short
\citep[e.g.,][]{postnov2015,sheng2023}.
From the phase-dependent spectral flux,
we derive observables such as pulse profiles,
phase-averaged
spectra, and pulsed fraction spectra.

\subsubsection{Polarisation}

We calculated the linear polarisation of both the
internal and observed emission using the mode
intensities $I_{1,2}$. In the neutron-star rest frame,
we choose a local basis such that the sign of Stokes
$Q$ distinguishes polarisation parallel and
perpendicular to the magnetic field,
\begin{equation}
\label{eq:qstock}
    Q_\mathrm{loc} = p_{q,1}I_1 + p_{q,2}I_2\,,
\end{equation}
where $p_{q,j}=p_{q,j}(E,\theta)$ are the signed
linear polarisation fractions \citep{pavlov1979, meszaros1992}.
Under the conditions studied, the modes remain largely
orthogonal, $p_{q,1}=-p_{q,2}$, but their values can
deviate substantially from unity depending strongly on
$\theta$. After applying gravitational corrections to the
intensities, we rotated $Q_\mathrm{loc}$ into the
observer's frame using the position angle, $\chi$,
of the projected magnetic field
\citep[as prescribed by the rotating vector model,
RVM; see, e.g.,][]{meszaros1988a, poutanen2024},
obtaining $Q_\mathrm{obs}$ and $U_\mathrm{obs}$.
Here we performed these calculations only for hot
spots in the single-emitting-point approximation,
unsuitable for extended columns. Further
details are given in Appendix~\ref{sec:appa}.

\section{Internal emission: spectra and beaming}
\label{sec:res:sigelt}

\subsection{Polarised flux}
\label{sec:res:sigelt:spec}

\begin{figure*}
  \centering
  \includegraphics[width=\textwidth]{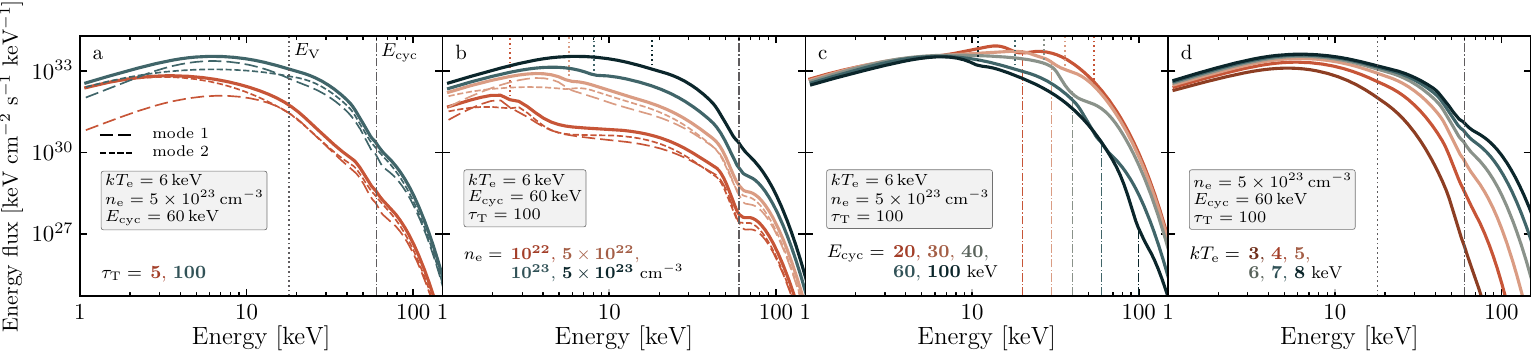}
  \caption{Energy flux in the neutron-star rest frame for
  different parameters combinations for the slab model
  (cylinder model is qualitatively similar; see Sect.~\ref{sec:res:sigelt:spec}).}
\label{fig:fflux}
\end{figure*}

The emergent flux in polarisation mode $j$, obtained by
integrating the specific intensity over angle in the rest
frame of the emitting plasma, $F_j(E)=2\pi\int_0^1 f_j(\mu, E)\,\mathrm{d}\mu$,
conveniently characterises the overall spectral shape.
The differential flux $f_p$  is $I_j(\mu, E)\mu$
for the slab and
$I_j(\mu, E)\sqrt{1 - \mu^2}$ for the cylinder (MN85a).
Figure~\ref{fig:fflux} shows how the principal parameters
of the \texttt{sigel-T} model affect the spectral flux in
the neutron star rest frame. The flux from a cylinder is
qualitatively similar to that from the slab, but has a more
pronounced cyclotron line. We therefore focused on the slab
here and in Sect.~\ref{sec:res:sigelt:cyc}. The main
difference lies in the emission angular distribution,
presented in Sects.~\ref{sec:res:sigelt:specf} and
\ref{sec:res:sigelt:beam}.

The overall spectral shape is largely determined by the
difference between the Comptonized spectra of the two
polarisation modes. For high-density media of intermediate
optical depth, $\tau_\T\sim5$--$20$, our results
qualitatively agree to those of
\citet{lyubarskii1987, lyubarskii1988} and
\citet{ceccobello2014}, who calculated the spectra below
the cyclotron resonance energy. In this regime, soft photons
are produced mainly in mode~2 and gain energy through
repeated scatterings, with a finite probability to
convert into mode~1 and escape the medium. As a result,
mode~2 dominates the low-energy part of the spectrum,
while photons converted to mode~1 contribute increasingly
towards higher energies (see, e.g., the case of $\tau_\T=5$
in Fig.~\ref{fig:fflux}a). At higher optical depths (e.g.,
$\tau_\T=100$), a sufficient number of photons is also
emitted in mode~1 deep in the medium. Their low cross
section allows them to escape from deeper layers,
producing a soft-energy excess
\citep[][see also Fig.~\ref{fig:crs}]{nagel1981b, sokolova-lapa2021}.

The complex spectral shape arises from the interplay
between the polarisation modes and redistribution near
the cyclotron and vacuum resonances (marked by dotted
and dash-dotted lines, respectively, in
Fig.~\ref{fig:fflux}). Several features appear as
artifacts of our model: a strong suppression just
above $E_\V$ at low electron densities due to the
assumption of a homogeneous medium
\citep{sokolova-lapa2023}, and the significant decrease
in spectral hardness with increasing $E_\cyc$, caused
by the lower emissivity. In a self-consistent treatment,
the latter would substantially increase the medium's
temperature because cooling becomes less efficient at
higher magnetic fields \citep{meszaros1983, harding1984}.

\begin{figure}
  \centering
  \includegraphics[width=0.95\columnwidth]{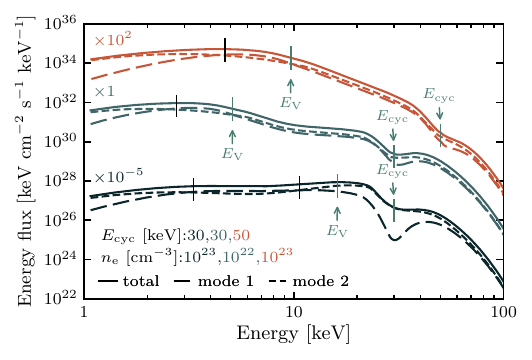}
  \caption{Energy flux in two modes for the slab model,
           shown for different $E_\cyc$ and $n_\el$;
           $kT_\el=5\,\mathrm{keV}$, $\tau_\T=20$. Vertical
           lines mark $E_\cyc$, $E_\V$, and mode-crossing
           points. The spectra are scaled for clarity,
           as indicated.}
\label{fig:fshape}
\end{figure}

\subsection{Cyclotron and vacuum resonances}
\label{sec:res:sigelt:cyc}

Depending on the physical parameters of the medium, several
characteristic energies can be identified by changes in
spectral shape or polarisation properties: $E_\V$, $E_\cyc$,
and the continuum mode crossing points, 
marked in Fig.~\ref{fig:fshape} for different $n_\el$
and $E_\mathrm{cyc}$.
The cyclotron line is typically dominated by mode~2
and is therefore sensitive to the relative plasma
and vacuum contributions to the photon polarisation vectors.
Lower electron densities and stronger magnetic fields
shift $E_\V$ to lower energies, placing more of the X-ray
spectrum in the nominally vacuum-dominated regime,
$E\gtrsim E_\V$ (Fig.~\ref{fig:crs} and Fig.~\ref{fig:fflux}b,c).
This classification
is very approximate, however, because neither contribution
vanishes abruptly at the vacuum resonance energy. Consequently, the
cyclotron-line core can remain strongly affected by the plasma
even when $E_\V<E_\cyc$. Resulting suppression of the mode~2 resonance
(Fig.~\ref{fig:crs}) produces a shallower line and modifies
the emission beaming. Both contributions also continue to
determine the ellipticities of the polarisation modes.

Across the total optical depths studied here, $\tau_\T\ge5$,
the angle-averaged cyclotron line profile changes little with
$\tau$. The exception is broadening and strengthening of
emission wings with the high $\tau_\T$ and with increasing
$kT_\el$. Hereafter, we call the parts of the absorption
profile outside the line core ``absorption flanks'', and
redistribution-affected continuum region around the line
``emission wings'', even when the latter produces no obvious
excess above the local continuum (this choice is more obvious
while considering the angle-resolved line profile; see
Sects.~\ref{sec:res:sigelt:specf} and \ref{sec:res:sigelt:beam}).
We refer to both of them generally as ``wings'' and note
that they can extend over tens of keV for the characteristic
$E_\cyc$ values, creating a more complex continuum shape.
Our exploratory fits with
phenomenological continua, such as a power law with a
high-energy cutoff, often required spurious shallow emission
or absorption features below the fundamental cyclotron line,
similar to the so-called 10\,keV feature
\citep[see also][for a similar discussion]{sokolova-lapa2023}.
The wings are weaker at stronger fields,
$E_\cyc\gtrsim60\kev$, over the chosen range of $kT_\el$,
as fewer photons populate the resonance for redistribution.

\subsection{Angular dependence of the mode spectra}
\label{sec:res:sigelt:specf}

\begin{figure*}
  \centering
  \includegraphics[width=0.92\textwidth]{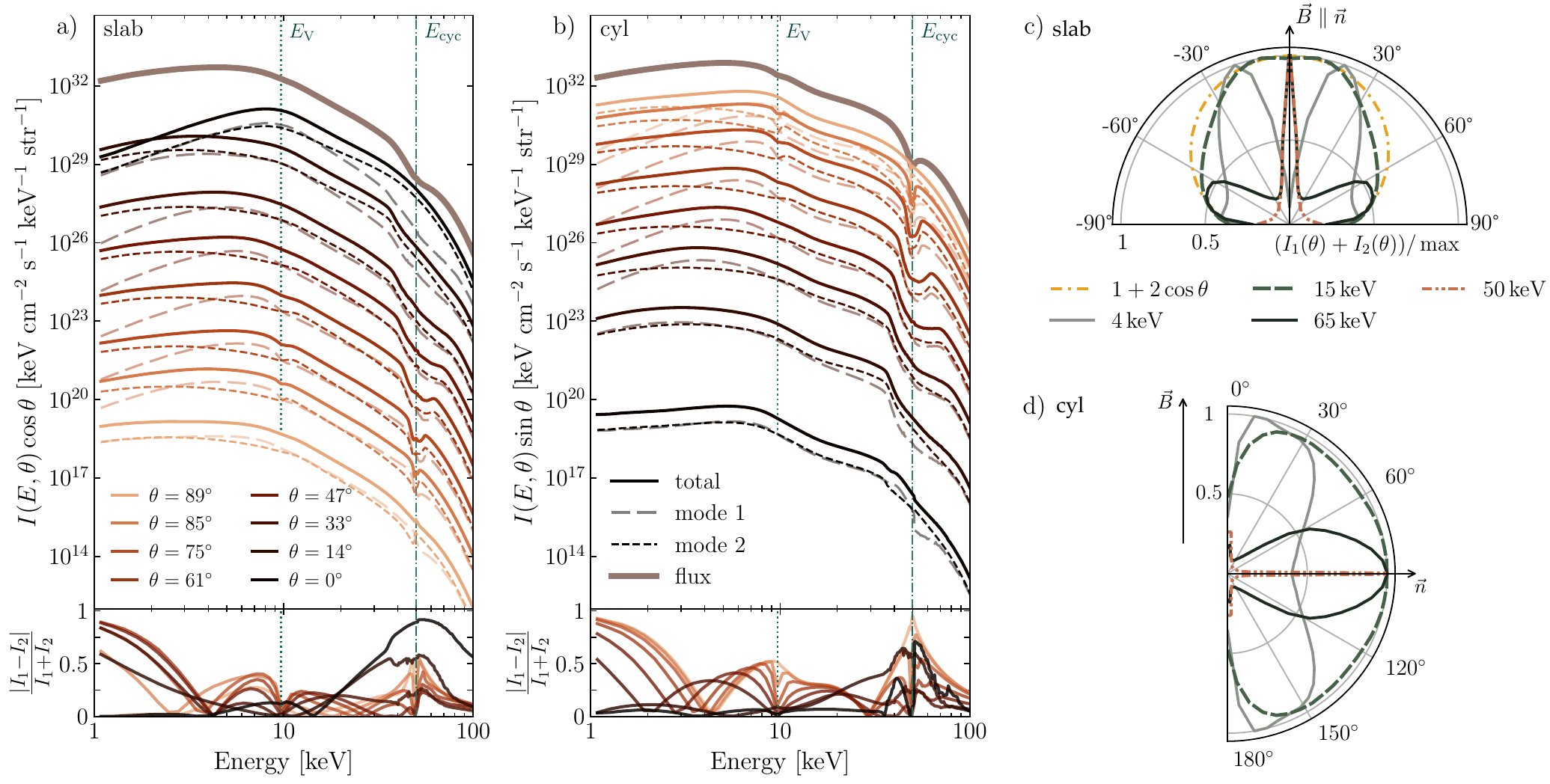}
  \caption{
  Angular dependence of the emission emerging from strongly
  magnetised plasma.
  Panels a and b show the differential flux in the two
  polarisation modes (top) and the corresponding mode-intensity
  difference (bottom) for the slab and cylinder, respectively.
  The thick upper line in each panel shows the total energy flux,
  $F=F_1 + F_2$. The thin lines show the differential flux at
  successive angles, starting from $\theta\sim0^\circ$
  for the slab and $\theta\sim90^\circ$ for the cylinder; the
  first is unscaled, and each subsequent spectrum is reduced
  by a factor of ten.
  The $0^\circ$ label denotes $\theta=0.2^\circ$. Panels c
  and~d show the corresponding beam patterns at different
  energies, summed over the two polarisation modes, together
  with the commonly adopted phenomenological profile. The
  reference model parameters are
  $E_\cyc=50\kev$, $n_\el=10^{23}\,\mathrm{cm}^{-3}$,
  $kT_\el=5\kev$, and $\tau=20$ and 50 for the slab and
  cylinder, respectively.
  }
\label{fig:emprof}
\end{figure*}

Thermal Comptonization in a strong magnetic field
produces anisotropic emission in continuum and the
cyclotron line. Figure~\ref{fig:emprof} shows the
differential spectral flux at different $\theta$
for the models hereafter referred to as the reference
slab and cylinder models. Both have $E_\cyc=50\kev$, $n_\el=10^{23}\,\mathrm{cm}^{-3}$,
$kT_\el=5\kev$, with $\tau=20$ and $50$,
respectively\footnote{The difference in optical
depth compensates for more efficient photon escape
from a cylindrical region (MN85b), placing the two
reference models in approximately similar transfer regimes.}. Appendix~\ref{sec:appd} presents several models for
other parameter combinations. Away from
the field direction, at $\theta\gtrsim15^\circ$, the
continuum shape depends only mildly on $\theta$,
unlike the relative mode contributions and the
vacuum- and cyclotron-resonance features. The latter
broadly traces the angular behaviour of the scattering
cross sections (Fig.~\ref{fig:crs}).
However, resonant redistribution of photons in energy
and angle during radiative transfer through the
optically thick plasma, modifies line profiles, in
particular its width, beyond what can be inferred from
the total cross sections alone. Previous studies
related the angular variation of the line with to
the Doppler width for Maxwellian electrons
$w_\mathrm{G}=E_\cyc\sqrt{kT_\el/m_\el c^2}
|\cos\theta|$ \citep[e.g., MN85a,][]{schonherr2007}.
To quantify it, we
fitted spectra shown in Fig.~\ref{fig:emprof}
above 10\,keV with a cutoff power-law continuum
with a Gaussian absorption line (see Appendix~\ref{sec:appc}).
The emergent cyclotron lines are substantially broader,
and their widths increase towards the magnetic field
much more steeply than predicted by Doppler brodening
(see Fig.~\ref{fig:sigma}).

The additional broadening is a transfer
effect caused by resonant redistribution, rather
than a property of the microscopic resonance profile
itself. At smaller angles, the resonance remains
optically thick over a wider energy range, so photons
undergo repeated resonant scatterings and diffuse
farther in energy before escaping.
Consequently, the emergent cyclotron line is
significantly broader than the resonance in
the total scattering cross section (Fig.~\ref{fig:sigma}). Interpreting
its width in terms of thermal Doppler broadening
would overestimate the plasma temperature by factors
of several \citep[see also][for the related discussion]{araya1999}.

The vacuum resonance produces a line-like
depression only at angles $\theta\gtrsim40^\circ$.
At smaller angles, the polarisation modes evolve
gradually across $E_\V$, and the spectrum is dominated by
continuum Comptonization in the individual modes.
Although the mode difference remains small near
$E_\V$, it does not necessarily vanish because
photons are redistributed across the resonance energy.
The detailed behaviour in this region should, however,
be interpreted cautiously, as it requires treatment
of the transfer for all four Stokes parameters,
allowing for partial mode conversion
\citep{van_adelsberg2006, garasev2016}.

In the normalised mode-intensity difference, the
continuum crossing produces a feature
similar to the vacuum resonance. Depending on the plasma
parameters, both may occur below
${\sim}10\,\mathrm{keV}$ and shift in
energy with the angle to the magnetic field. The
mode-intensity difference is not, however,
a reliable proxy for the linear PD
(Eq.~\ref{eq:qstock}) as at
$\theta\lesssim30^\circ$, their propagation is mainly
determined by the plasma response and thus the modes
are elliptically or circularly polarised (Sect.~\ref{sec:res:pol}).

\subsection{Shapes of fan and pencil beams}
\label{sec:res:sigelt:beam}

Figure~\ref{fig:emprof}(c,d) show shapes and the
energy evolution of the beam patterns, $I(\theta)$ at
four representative energies, $4\kev$, $15\kev$,
$E_\cyc$, and $1.3E_\cyc$, sampling the soft and
intermediate-energy continuum bands, cyclotron line,
and the blue wing
In the slab geometry, the
intermediate-energy beam is featureless and pencil-like,
with angular pattern broadly described by
$1+b\cos^n\theta$, where $n=1$--$3$ and $b>0$ depend
on the model parameters (see Appendix~\ref{sec:appe}
for more examples and also \citealt{leahy1990} for
power-series fits in $\cos^2\theta$ to
$I(\theta)\cos\theta$ for a similar MN85a,b model).

At soft energies, the beams commonly have a more
complex shape due to a pronounced split along the
magnetic field direction. This shape arises for
photons below ${\sim}10\kev$ because the scattering
optical depth crosses unity with increasing angle to
the magnetic field, taking the slab from optically thin
near $\theta\sim0^\circ$ to optically thick at larger
angles (\citealt{basko1975}, MN85b).
In the phenomenological profile, this split can be
empirically represented by adding a multiplicative
factor 
$\left[1-(1-f_0)\exp\left(-\frac{\theta^2}{2\eta^2}\right)\right]$,
with $f_0\approx0.2$--$0.7$, and $\eta\approx5^\circ$--$15^\circ$.
The split deepens and widens, while the intermediate-energy
beam broadens, with increasing $E_\cyc$, decreasing
$\tau_\T$ and $n_\el$, and, more weakly,
increasing $kT_\el$ (see also Fig.~\ref{fig:ppem_spot}).
Under the opposite conditions, the split may disappear
altogether.

Across the broad energy range around $E_\cyc$,
${\sim}0.7$--$1.5E_\cyc$, the pattern
develops a narrow, sharp central beam aligned with
the field direction and two side petals. The petals
are weak and peak at ${\sim}90^\circ$ at $E_\cyc$,
but strengthen and shift towards ${\sim}70^\circ$
in the line wings. The petals reflect the angle-dependent
line profile: in Fig.~\ref{fig:emprof}, at $1.3E_\cyc=65\,\mathrm{keV}$,
the emission pattern samples the absorption
blue flunk at $\theta\lesssim60^\circ$ but the 
emission wing or the continuum at larger angles,
producing a rapid intensity rise. It also reflects
that near $E_\cyc$, the larger angles are
more transparent for photons (MN85b).
The same effect occurs in the red wing (see, e.g.,
Fig.~\ref{fig:vacpl}a). The central beam is due to
the absence of the line in the spectra at $\theta\sim0^\circ$,
caused by the plasma-induced suppression of the
resonance in mode~2, which mostly dominates the line
core. The beam broadens with decreasing $E_\cyc$ and
increasing $n_\el$, that is, with the increasing plasma
influence on photon propagation. In the pure vacuum case,
strong resonances are imprinted even at $\theta=0^\circ$
and the central beam disappears (see Fig.~\ref{fig:emprof_plvac}).
By contrast, the increasing influence of the vacuum
effects narrows the central beam and produces the
side petals (Fig.~\ref{fig:vacpl}).

In the cylinder case, emission through the
walls forms a fan beam at intermediate energies,
which is much broader than the phenomenological
profile (Fig.~\ref{fig:emprof}d)\footnote{
A $90^\circ$-rotated $(1+b\cos\theta)$ beam is commonly
adopted for intrinsic column-wall emission, with
later downward boosting by the bulk flow \citep[e.g.,][and references therein]{poutanen2013,markozov2024,falkner2026a,falkner2026b}.}.
At soft energies, emission is often suppressed along
the wall normal, $\theta\sim90^\circ$, where mode~2
dominates. The effect is created by the increasing
mode~1 contribution at intermediate angles. A minimum
at ${\sim}0^\circ$ occurs in both profiles throughout
the parameter grid. When this direction is optically
thin for scattering, its origin is the same as for the slab;
otherwise, when thick, it follows the
modified Eddington-Barbier relation,
$I(\theta)\propto\sin\theta/\sigma(\theta)$ \citep[][]{nagel1981a}.
Near the cyclotron resonance, the pattern again
develops a central beam, now perpendicular to the
field, along the surface normal. Much broader than
in the slab, it is flanked by two narrow side petals,
only a few degrees wide, extending towards the field
direction.

\section{Observed flux}
\label{sec:res:lbsigelt}

\subsection{Pulse profiles}
\label{sec:res:lbsigelt:pp}

\begin{figure*}
    \centering
    \includegraphics[width=0.95\textwidth]{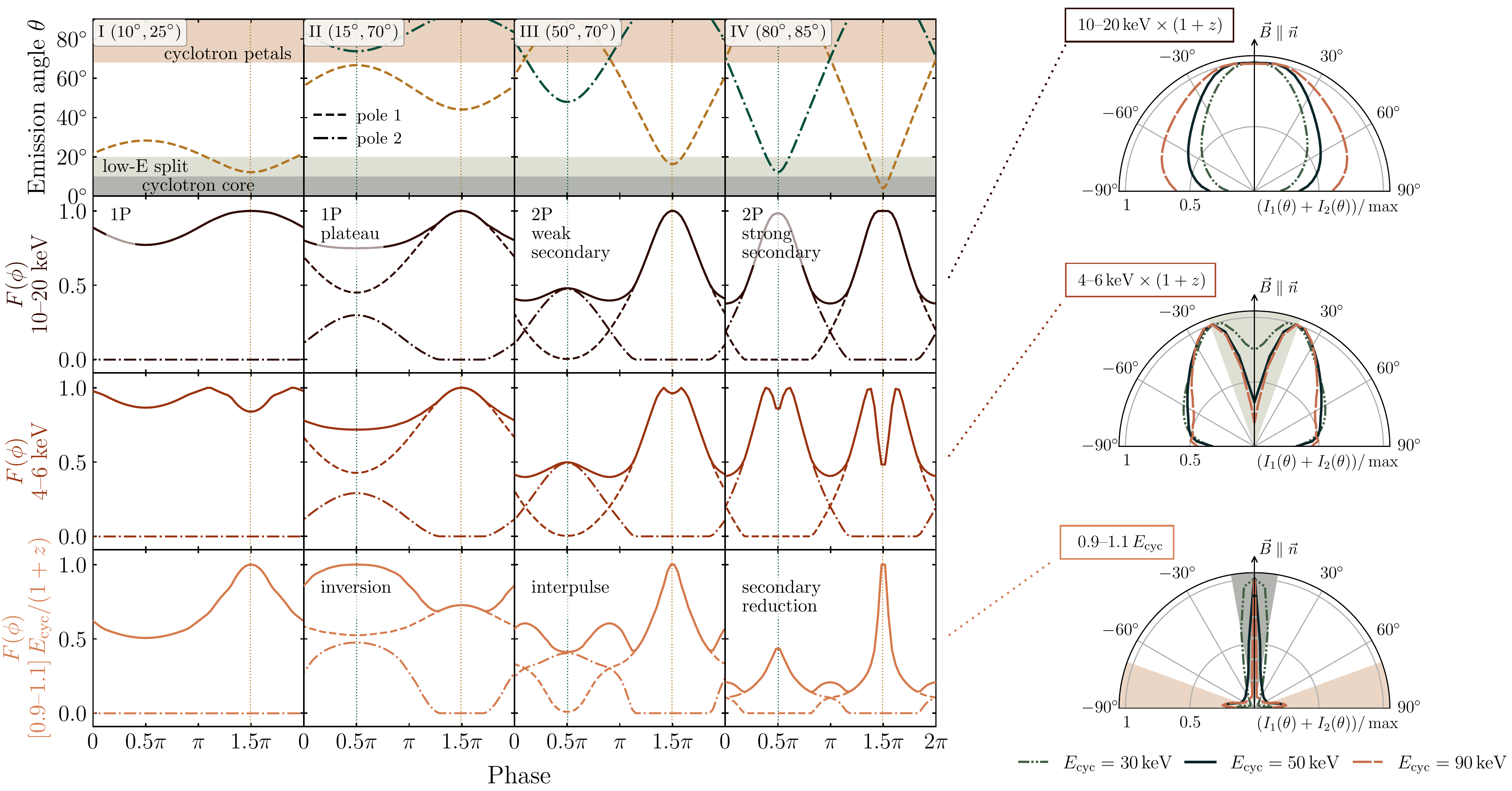}
    \caption{Phase variation of the emission angles of radiation
    reaching the observer from the centres of two hot spots
    $r_0=140\,\mathrm{m}$, calculated including light bending, for
    four geometries, $(i_1, i_2)$, and the corresponding pulse
    profiles in selected energy bands; $\mathcal{C}=0.34$
    ($z\approx0.23$). The internal-emission parameters are as
    in Fig~\ref{fig:emprof}. The right-hand panels show
    band-integrated beam patterns for the reference model and
    variants with $E_\cyc=30\,\mathrm{keV}$ and $E_\cyc=90\,\mathrm{keV}$.
    Shading in the beam-pattern and emission-angle panels marks
    the angular ranges of selected features, including the central
    split at soft energies and the cyclotron-beam core and petals.}
    \label{fig:ppem_spot}
\end{figure*}

\begin{figure*}
    \centering
    \includegraphics[width=0.95\textwidth]{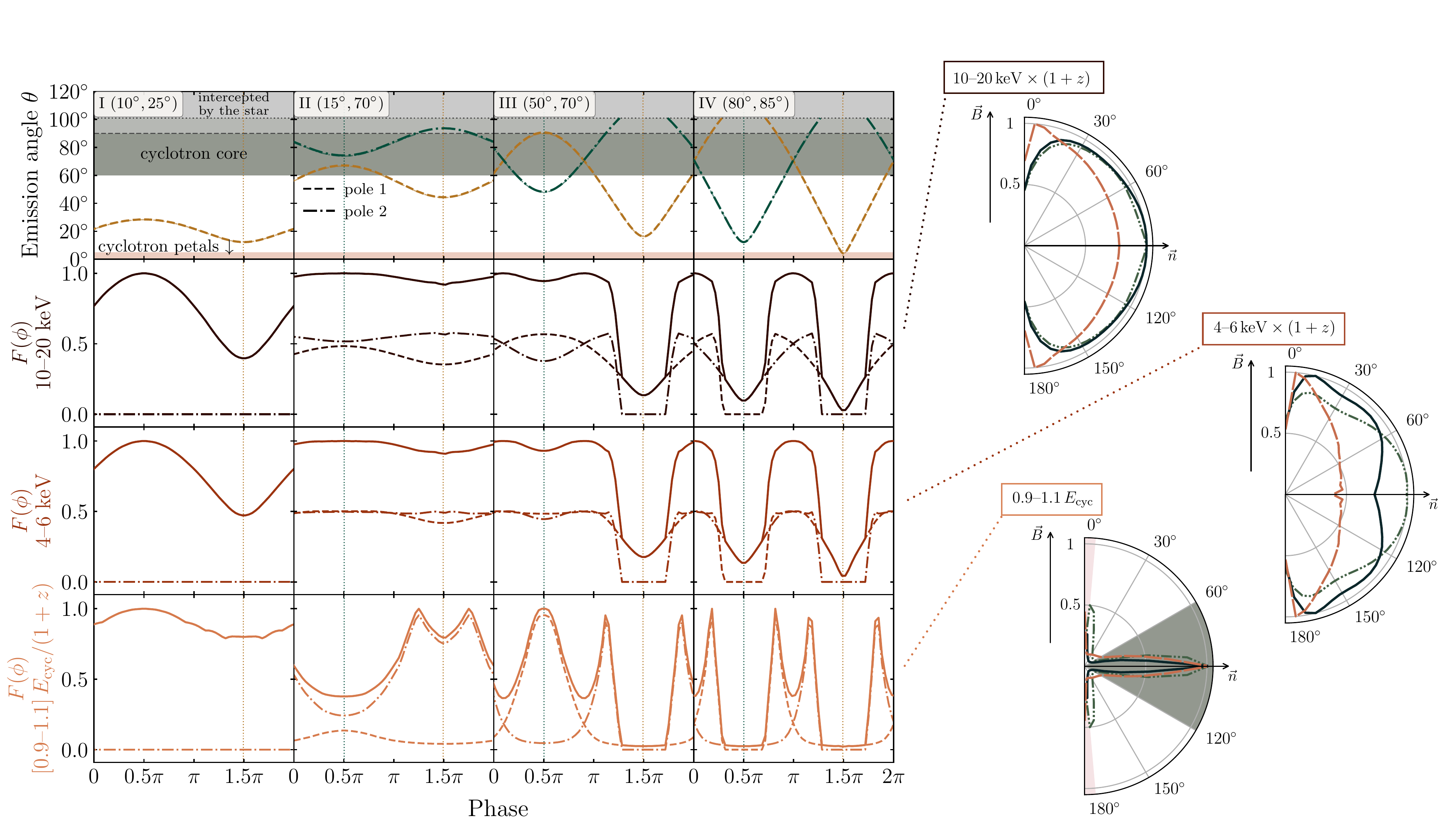}
    \caption{Same as in Fig.~\ref{fig:ppem_spot} but for accretion
    columns with $h=300\,\mathrm{m}$. Grey densely dashed and dotted
    horizontal lines mark the emission angles at which the column
    base and top, respectively, are occulted. The parameters of
    the three models shown in the right-hand beam panels are the
    same as in Fig.~\ref{fig:ppem_spot}.
    }
    \label{fig:ppem_cyl}
\end{figure*}

Following Sect.~\ref{sec:res:sigelt:beam}, we considered
pulse profiles in the soft, $4$--$8\kev$, intermediate,
$10$--$20\kev$, and cyclotron-line bands. To account for
gravitational redshift, we adopted $[0.9E_\cyc, 1.1E_\cyc]/(1+z)$, for the latter, where
$1+z=1/\sqrt{1-\mathcal{C}}$ is the surface redshift
factor\footnote{Column emission combines contributions
from different heights and hence redshifts, i.e., different
gravitational redshifts. For the low columns considered here,
however, even the maximum height, $h=300\,\mathrm{m}$, changes
the energy shift, $(1+z)^{-1}$, by only 0.4--1\% from its
surface value across the adopted range of $\mathcal{C}$.}.
Given the strong anisotropy, it is instructive to identify
which emission angles, and hence which parts of the beam pattern,
are visible at each phase. For the spatially compact
emission regions considered
(Sect.~\ref{sec:met:lbsigelt:emang}), we track the internal
emission angle at the centre of the region, accounting for
light bending. Figure~\ref{fig:ppem_spot} shows its phase
evolution for hot spots with $r_0=140\,\mathrm{m}$ in four distinct geometries,
$(i_1, i_2)$, together with pulse profiles in the selected
energy bands and the corresponding band-integrated beam
patterns.

For emitting hot spots and intermediate energies, the
relative contribution of the primary and secondary poles
determines the pulse profile morphology, changing from
single- to double-peaked as the secondary pole
contribution increases. Each peak occurs near the
closest approach of the corresponding pole to the observer.
Because the beam at this band closely resembles the
commonly adopted pencil-beam approximations, the
resulting pulse profiles follow a similar geometric
trends found for $\cos\theta$ beaming \citep{annala2010}
and, more generally, for isotropic emission
\citep{beloborodov2002}. Following these works, in
Fig.~\ref{fig:ppem_spot} we suggest a classification
of the profiles as one- or two-peaked (``1P/2P''),
with further distinctions based on the secondary-peak
strength.
The general pulse-profile morphology for geometries
remains unchanged across the explored parameter ranges,
including the broad range of explored $E_\cyc$,
consistent with the stability of the beam patterns.
Factors that broaden the beam, such as a higher magnetic
field (Sect.~\ref{sec:res:sigelt:beam}), also broaden
the pulse profiles. Increasing the polar cap radius
to $500\,\mathrm{m}$ does significantly alters the
profiles.

The low-energy pulse profiles follow a similar trend.
However, for a broad range of physical parameters, the
beam is split along the magnetic-field direction,
producing a central dip in the pulse-profile peak
whenever the minimum emission angle of the corresponding
pole falls below ${\sim}10$--$20^\circ$. Such a feature
therefore indicates that the pole is viewed at a small
emission angle during its closest approach to the observer
and also provide a very general view onto physical
conditions in the emission-forming regions, as the split
becomes increasingly stronger with $E_\cyc$, lower $n_\el$
and $\tau_\T$ (see band-averaged beams in Fig.~\ref{fig:ppem_spot}).

Because of the complex shape of the cyclotron-line beam,
pulse profiles at these energies can exhibit markedly
different morphologies depending on geometry.
If the visible angles never reach the side petals
(see marked regions in Fig.~\ref{fig:ppem_spot}),
the cyclotron-line pulse profiles resemble the
intermediate-energy counterpart, with
each maximum corresponding to the minimum emission
angle of a pole (geometry I). However, the narrow
central beam produces sharper peaks, a weaker
secondary-pole contribution, and stronger modulation,
particularly when its most intense part is sampled.
If one pole traverses the petal-dominated angular
range, the petals create an interpulse (geometry~III).
If both poles are visible only at large angles,
$\theta\gtrsim50^\circ$, their petal emission
combines into a strong secondary peak, inverting
the profile relative to that at intermediate energies.
For pole-symmetric configurations, this is
the only mechanism producing a pronounced phase shift
between the intermediate-energy and cyclotron-line
profiles, of half a rotational cycle. In
geometries with interpulses, asymmetry
between the poles can instead cause the main peak
and an interpulse to merge, shifting the profile
maximum by ${\sim}0.1$--$0.5$ phases.
Increasing the magnetic field strengthens the side
petals of the cyclotron beam, producing stronger
interpulses and earlier transition to profile
inversion. The only exception to the trend in
Fig.~\ref{fig:ppem_spot}, occurs for models with
$E_\cyc=20\,\mathrm{keV}$, where the increasing
plasma contribution broadens the central cyclotron
beam, making the cyclotron-line profiles resemble
the intermediate-energy ones. At such low fields,
the effects of higher harmonics, which are not
included in our model, become important and should
be considered before drawing firm conclusions.

Figure~\ref{fig:ppem_cyl} presents the emission-angle
phase evolution, pulse profiles, and beam patterns for
columns with $h=300\,\mathrm{m}$. Unlike hot spots,
columns contribute the least when their poles are closest
to the observer because their tops do not emit in our model
and all beam patterns weaken towards the magnetic-field axis.
Moreover, the low columns considered here are fully occulted
when passing behind the neutron star and thus produce
no Einstein-ring peak \citep{meszaros1988b}. Occultations
cause sharp flux drops, only weakly filled by the
front column, viewed close to its axis.
Grey densely dashed and dotted lines in the top panel
of Fig.~\ref{fig:ppem_cyl} mark the maximum visible
emission angles at the column base and top, respectively.
The interval between them indicates partial occultation,
with the base hidden but the top visible
\citep[see Fig.~2 in][]{falkner2026b}.

Apart from minor differences in modulation, the low-
and intermediate-energy profiles have similar
morphologies. Because the two poles generally contribute
most strongly at similar phases, most geometries produce
single-peaked profiles. Double peaks arise when both
geometric angles are large enough for both columns to be
occulted (e.g., IV in Fig.~\ref{fig:ppem_cyl}). The
resulting flux rises and drops are much sharper than for
an emitting hot spot. At the cyclotron line, the narrower
beam produces even steeper flanks and confines each pole’s
contribution to a smaller phase interval.
If a pole remains visible throughout the cycle but
is never viewed at angles sampling the central beam,
${\sim}60^\circ$--$80^\circ$, its contribution
remains very weak. Otherwise, its flux increases
with emission angle until the column becomes
partially or fully occulted.
For the column, varying the physical parameters
within $E_\cyc=30$--$100\,\mathrm{keV}$ does not alter
the overall pulse-profile morphology across geometries
for all three energy bands. Again, the lowest-field
models, $E_\cyc=20\,\mathrm{keV}$, is the exception:
its cyclotron-line pulse profiles broadly resemble
those in the other two energy bands. Decreasing the
column height mainly narrows the peaks without
changing their morphology, except in geometry~II
at phase $1.5\pi$, where the shortest column
($h=10\,\mathrm{m}$) is fully occulted and its
flux drops to zero.

\subsection{Pulsed fraction near cyclotron line}
\label{sec:res:lbsigelt:pfs}

\begin{figure*}
  \centering
  \includegraphics[width=0.9\columnwidth]{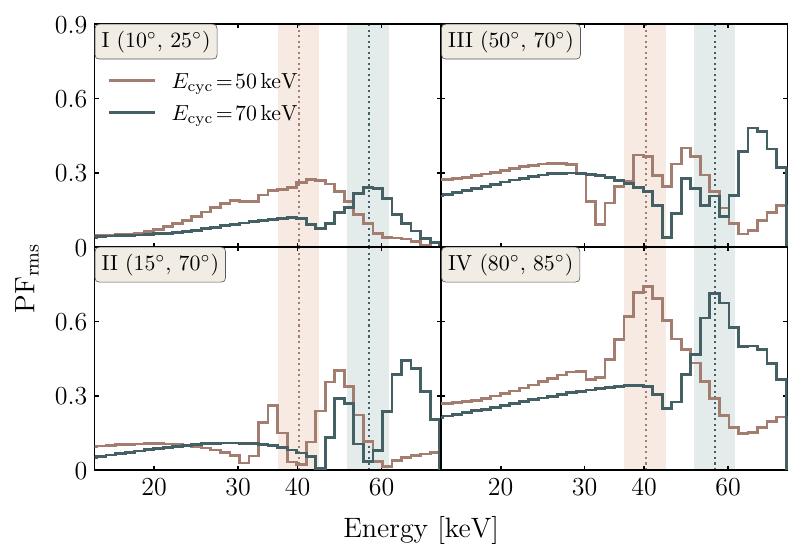}
  \hspace{1.5em}
  \includegraphics[width=0.9\columnwidth]{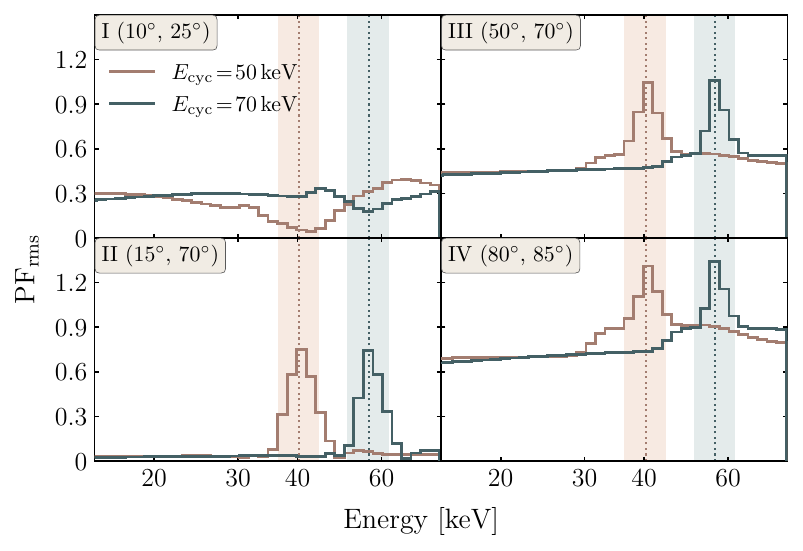}
  \caption{Pulsed fraction spectra for hot spots (left)
  and accretion columns (right) for two $E_\cyc$ (the
  reference model and a variant with $E_\cyc=70\,\mathrm{keV}$).
  Dotted lines mark the redshifted cyclotron energies, and
  shaded regions the corresponding intervals, $0.9$--$1.1E_\cyc$.}
\label{fig:pfs}
\end{figure*}
The pronounced change in pulse-profile morphology
between the intermediate and cyclotron-line bands
is expected to leave a characteristic imprint on
the pulsed fraction spectrum. Here,
we adopt the root-mean-square (rms) definition of
the pulsed fraction \citep[see][for details]{ferrigno2023,pottschmidt2026}.
Figure~\ref{fig:pfs} shows the pulsed fraction
spectra for the geometries I-IV as before, for $E_\cyc=50\kev$ and $70\kev$.
For spot emission (left panel), the
behaviour near the cyclotron line is again governed by
whether the LOS samples the beam core
or its petals. The pulsed fraction increases,
producing a bump across the cyclotron-line
profile in two cases: when the petals
are not visible ($i_1,i_2\lesssim40^\circ$, e.g., I)
or when the LOS also passes close to the
beam axis, so that the brightest part of
central beam dominates the pulsed fraction
($\Delta i=|i_1-i_2|\lesssim5$--$10^\circ$, e.g., IV).

The contribution of the side petals relative to
the core increases when the difference between
the two angles reaches $\Delta i\gtrsim10^\circ$,
provided that both angles remain $\gtrsim40^\circ$
such that both poles are visible (e.g., geometry
III). Because the
petals are more prominent in the red and blue
wings of the line, the pulsed fraction first
decreases in the wing energies. In the pulse profiles, this
corresponds to stronger interpulses in the wing energies
than in the line core. The resulting wing minima
produce a W-shaped pulsed fraction
spectrum in the broad vicinity of the cyclotron
line, as for geometry~III with $E_\cyc=50\kev$.
At higher magnetic fields, the central beam
narrows while the side petals strengthen.
Consequently, for $E_\cyc\gtrsim70\kev$, local
pulsed-fraction minima occur in the line wings
around a central bump even for geometries
with $i_1=i_2$.

A further increase in $\Delta i$ drives a 
transition from an interpulse to an inversion of
the cyclotron-line pulse profiles (Fig.~\ref{fig:ppem_spot}).
The growing contribution of the side petals in
the line wings produces inverted pulse profiles
with larger amplitudes than in the line core.
This initially creates a dip in the pulsed fraction
spectrum, often with complex substructure (see
geometry~III for $E_\cyc=70\kev$). With increasing
$\Delta i$, it develops into an inverted M-shaped
structure with pronounced maxima in the line wings.

For column emission, the stronger modulation of
the cyclotron-line pulse profiles caused by their
more restricted visibility (Sect.~\ref{sec:res:lbsigelt:pp})
produces a pulsed-fraction bump for most geometries
(II--IV). The exception is the small-angle geometry~I,
where only one pole is visible and is sampled from
angles away from the central beam. This produces nearly
flat cyclotron-line pulse profiles and hence
a dip in the pulsed fraction spectrum.

\subsection{Phase-averaged spectra}
\label{sec:res:lbsigelt:flux}

\begin{figure*}
  \centering
  \includegraphics[width=\columnwidth]{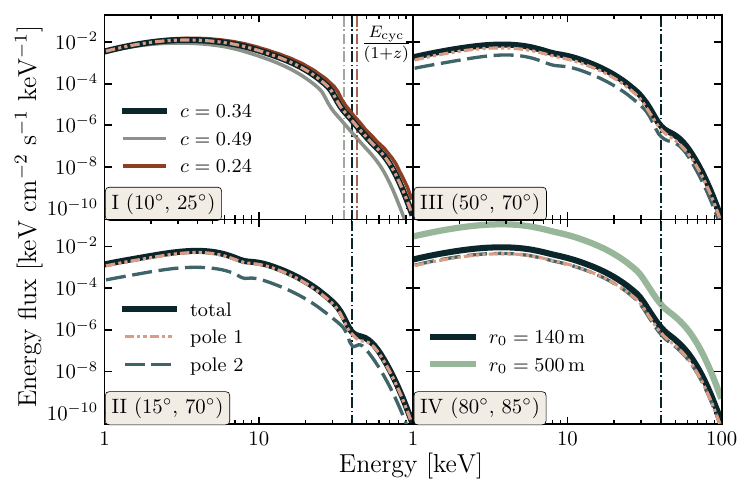}
  \includegraphics[width=\columnwidth]{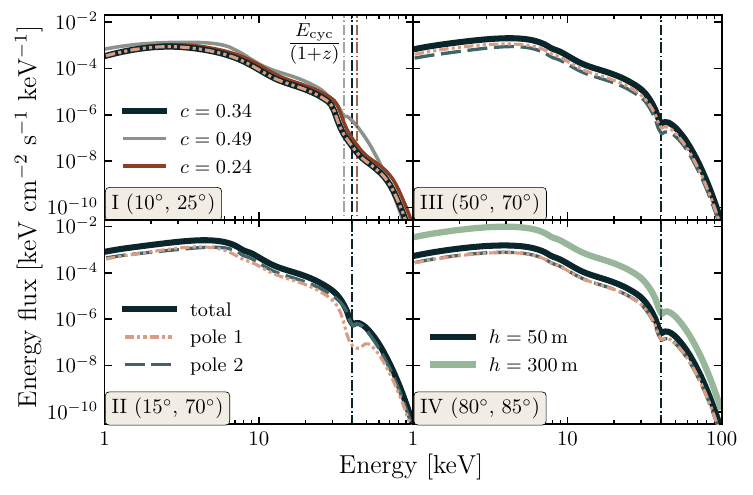}
  \caption{Phase-averaged spectra for hot spots (left) and
  accretion columns (right), including the individual pole
  contributions. Panels I and IV additionally show the total
  spectra for different compactnesses and emission-region extents,
  respectively.}
\label{fig:phav}
\end{figure*}

Figure~\ref{fig:phav}
presents phase-averaged spectra for spot-like (left)
and column-like emission regions (right), together with
the individual pole contributions. The overall
spectral shape, particularly at soft X-ray
energies, is relatively insensitive to geometry.
For a hot spot, the spectrum is dominated by the
primary pole for all compactnesses and spot sizes
considered here, except when either $i_1$
or $i_2=90^\circ$, in which case the two poles
contribute almost equally over most of the energy
range. Their relative contributions nevertheless
vary across the spectrum. For geometries in which
at least one of the angles is sufficiently large
for the secondary pole to become visible, its
relative contribution increases in the cyclotron-line
wings. The secondary pole is seen predominantly at
large angles to the magnetic field, $\gtrsim40^\circ$,
and therefore produces a narrower line. Its contribution
thus tends to reduce the width of the total cyclotron
feature (e.g., geometries II and III
in Fig.~\ref{fig:phav}). Only when both
$i_1,\,i_2\approx90^\circ$ the secondary pole exceed
the primary flux and dominates the cyclotron-line
region.

\begin{figure}
  \centering
  \includegraphics[width=0.95\columnwidth]{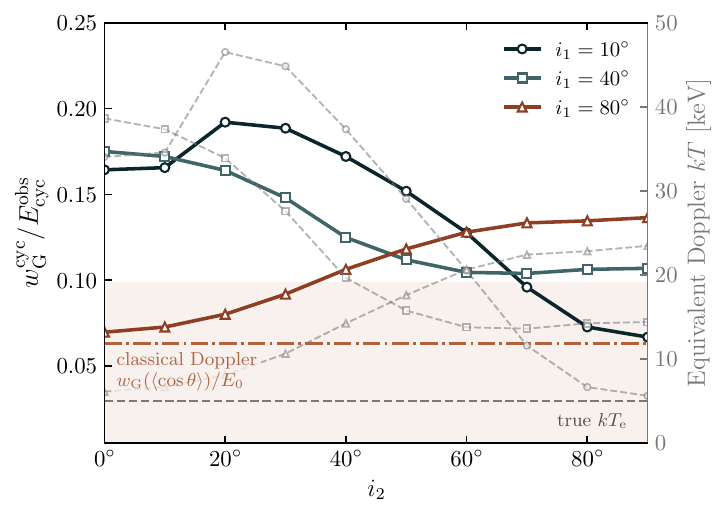}
  \caption{Geometry dependence of the fitted 
  cyclotron-line Gaussian width-to-centroid (broadening)
  ratio, $w^\mathrm{cyc}_\mathrm{G}/E^\mathrm{obs}_\mathrm{cyc}$,
  for hot-spot emission (reference model). The
  shaded region shows the Doppler-profile prediction
  for $\cos\theta\in[0,1]$. The right-hand axis displays
  the equivalent Doppler temperature, obtained as
 $m_\el c^2 \langle\cos\theta\rangle^{-2} (w^\mathrm{cyc}_\mathrm{G}/E^\mathrm{obs}_\mathrm{cyc})^2$.}
\label{fig:sigma_phav}
\end{figure}

The width of the cyclotron line in a spectrum
spans across a large range with geometry.
Similarly to Sect~\ref{sec:res:sigelt:cyc}, to
characterise it, we fitted the phase-averaged
spectra for various geometries for a phenomenological
model.
Figure~\ref{fig:sigma_phav} shows the resulting
Gaussian width of the line as a function of $i_2$
for three fixed values of $i_1$.
Although the line width varies significantly,
its variation is smaller than in the internal
emission across different emission angles
(see broadening ratios in Fig.~\ref{fig:sigma},
mainly driven by the width $w^\mathrm{cyc}_\mathrm{G}$
as the observed centroid energy,
$E^\mathrm{obs}_\mathrm{cyc}$ varies within
${\sim}10\%$).
This is primarily due to the $\cos\theta$
projection factor in the differential flux,
which suppresses contributions from large
angles, where the line is narrow, keeping the
integrated line fairly broad. The temperature
obtained by inverting the standard one-dimensional
Doppler-width relation is shown
on the right axis of Fig.~\ref{fig:sigma_phav},
illustrating that this simple relation can yield
values
far from the actual plasma temperature.

In the column case, a rapid switch of the dominant
pole occurs for geometries where $i_1$ or
$i_2\gtrsim20^\circ$. This switch first affects
the cyclotron line region, while keeping the
low-energy continuum dominated by the primary
pole (e.g., geometry II in Fig.~\ref{fig:phav}, right).
While the primary pole defines the cyclotron
line shape at small-angle geometries (I), its
profiles remains deep and broad. A transition
to the secondary pole dominating this energy
dramatically narrows the line shapes, as it
is now defined by the back column seen under
large angles to the magnetic field
\citep[similar to][but without
excessive emission wings]{falkner2026b}.
Both the high-energy contribution from the
secondary pole and the $\sin\theta$ projection
factor in the differential flux enhance
large-angle emission, producing a much narrower
line  and more pronounced emission wings than
in hot-spot emission for most geometries. The
fitted line width can be even smaller than the
corresponding classical Doppler width, causing
the temperature inferred in this way to
underestimate the actual temperature.
This effect becomes less pronounced at lower
magnetic-field strengths (Fig.~\ref{fig:sigma_phav_cyl}).

The size of the spot primarily increases the overall
flux level. By contrast, higher compactness
reduces the flux, and is accompanied by spectral
softening and an increased redshift of the
cyclotron line. For a column, increasing
the column height or a compactness in addition
strongly affect the visibility of the second
column. This makes the narrow cyclotron line
defined by the back column appear in the phase-averaged
spectra at smaller-angle geometries (see, e.g., 
the transition with compactness displayed for
the geometry~I in Fig.~\ref{fig:phav}, right).

\section{Predictions for soft X-ray polarisation}
\label{sec:res:pol}

\begin{figure*}
  \centering
  \includegraphics[width=1.03\textwidth]{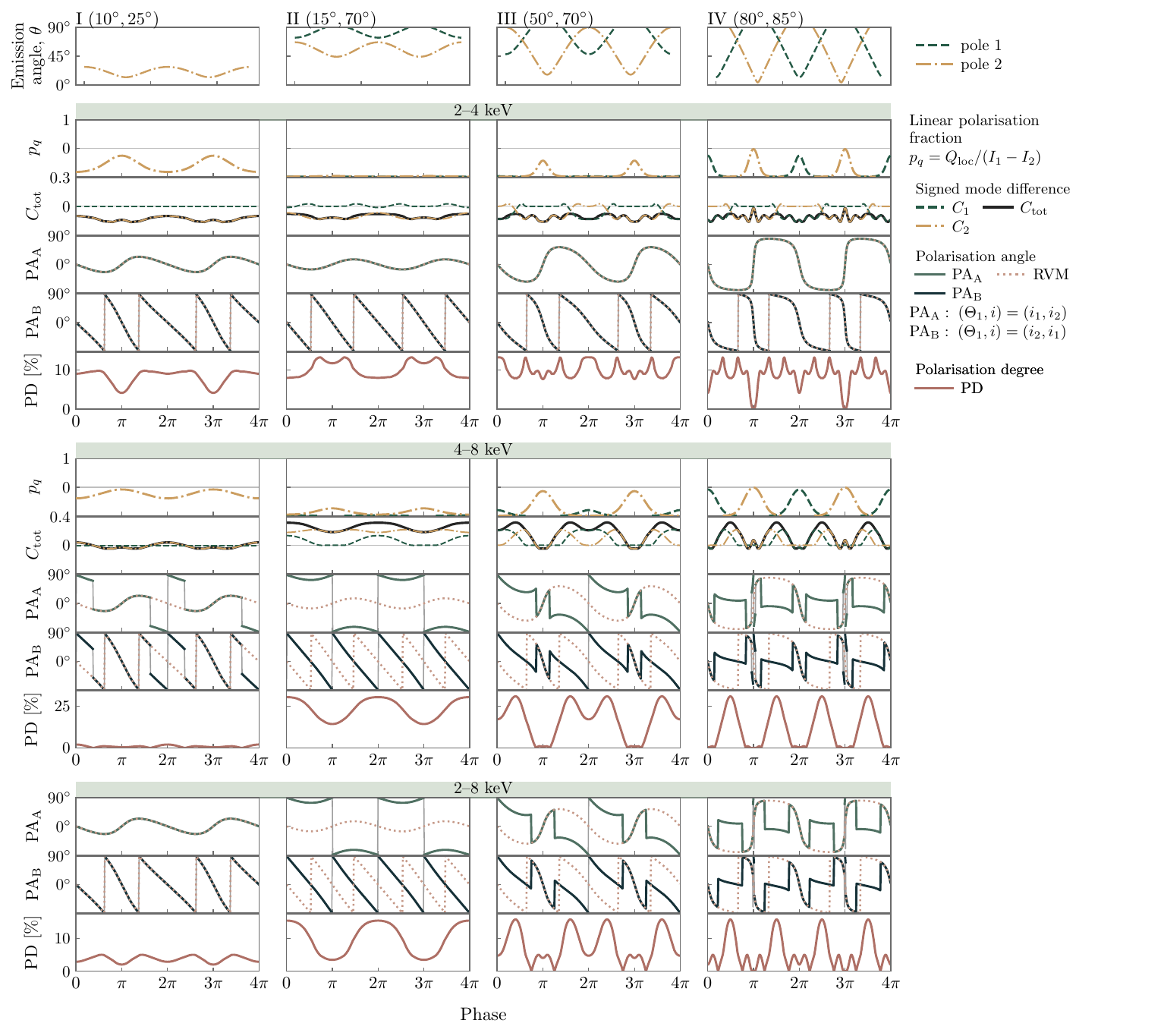}
  \caption{
  Phase dependence of band-integrated PA and PD for
  hot-spot emission in the reference model and geometries
  I--IV for there observed-energy bands. The two narrower
  energy bands additionally show $p_q$ for mode~1 at the
  band midpoint and the pole contributions $C_1$ and
  $C_2$ to the band-integrated signed mode difference
  $C_{\rm tot}$. PA is shown for configurations A,
  $(\Theta_1,i)=(i_1,i_2)$, and B, $(\Theta_1,i)=(i_2,i_1)$.
  Dotted curves show the corresponding RVMs.
  }
\label{fig:polaris}
\end{figure*}
We calculated the observable linear polarisation, focusing
on the soft-X-ray band relevant to \textit{IXPE},
while also exploring the broadband energy dependence of the PD
(Fig.~\ref{fig:pden}).
Although significantly more sensitive to
the physical parameters of the emission regions than,
for example, pulse profile, its general behaviour can be
understood in terms four main contribution: the relative
intensities of the two polarisation modes from each emitting
region, their intrinsic degrees of linear polarisation, the
flux-weighted combination of emission from the two poles,
and the integration of the Stokes parameters over a finite
energy band. As for the spectral observables, the polarisation
signal also depends strongly on the range of emission angles
sampled over the rotational cycle.

Figure~\ref{fig:polaris} shows
the intrinsic and observable polarisation quantities for
geometries I--IV and hot-spot emission in different energy
bands for the reference model (see Sect.~\ref{sec:appa} for
computational details). The top panels show the
phase-dependent emission angles and, together with
Fig.~\ref{fig:ppem_spot}, also trace the corresponding
pulse-profile morphology, except for the central split
(cf. Sect.~\ref{sec:res:lbsigelt:pp}).
The top panel in each band presents the linear polarisation
fraction $p_q$ at
the midpoint of each energy band. We plot only mode~1,
having verified that $p_{q,1}=-p_{q,2}$ holds to high
accuracy throughout the explored parameter space. The
modes become increasingly more circular towards the
magnetic field direction, with the angular extent of
this effect determined primarily by $n_\el$ and
$E_\cyc$\footnote{Near the cyclotron resonance, 
$kT_\el$ also has a strong influence.}. For our reference
model, the modes begin to depart from full linearity
already below $\theta\sim0^\circ$\footnote{The effect
persists even in a strongly vacuum-dominated regime,
although over
a narrower angular range: $\theta\lesssim20^\circ$ for
$n_\el=10^{21}\,\mathrm{cm}^{-3}$ and $E_\cyc=60\kev$
\citep[see, e.g., Fig.~4.4 of][]{sokolova-lapa2023phd}.}.
This is the main cause of the reduced linear
PD at the closest approach of the
magnetic pole to the observer's LOS,
$\theta=|i_1-i_2|$, corresponding to the maximum of the
intermediate-energy pulse profile \citep[see
geometry~I and][]{meszaros1988a}.

Mode dominance can be characterised by the band-integrated
signed mode difference, $C_\mathrm{tot}$, together with
the contributions from the individual poles (see
Eqs.~\ref{eq:ck} and ~\ref{eq:ctot}). If $C_\mathrm{tot}<0$
throughout the cycle, mode~2 dominates the band-integrated
spectrum (e.g., for 2--4\,keV, geometry~I). Within the band,
however, the dominant mode may change because of a mode
crossing or vacuum resonance in an individual pole
spectrum, or because the emission becomes dominated by
the other pole in a different mode.

Across the explored parameter space, mode crossings and
vacuum resonances occur often within the selected energy
bands, although rarely both in the same band.
In our reference model, the mode-crossing energy in the
$2$--$4\kev$ band varies with emission angle, spanning
the band from its lower boundary to approximately
$3.5\kev$ (see Fig.~\ref{fig:emprof}, but with the
applied redshift of $1+z=1.24$). Its resulting phase
dependence can be largely washed out by band integration 
ue to the phase-dependent motion of the crossing energy,
leaving a weak substructure in $C_\mathrm{tot}$ and the
PD (e.g. III and IV). By contrast, when the other pole
becomes dominant while emitting mainly in the opposite
mode, the resulting mode switch produces deeper PD minima,
as seen near $\phi=\pi/2$ and $3\pi/4$ for geometry~IV.
In the $4$--$8\kev$ band, the dominant mode from a single
pole instead changes gradually towards the field direction,
causing global minima in PD, supported by decreasing $|p_q|$
(e.g., III). Although this band includes the vacuum
resonance near its upper edge, at $7.9\,\mathrm{keV}$,
its effect here is negligible. Models for other parameters
show that the vacuum resonance in the energy band indeed
can be hidden in the energy-integrated polarisation
observables. However, the resonances and mode crossings
may be well reflected in the phase-integrated
$\mathrm{PD}(E)$, depending on the geometry (see
Fig.~\ref{fig:pden}).

Integration over a finite energy band is therefore
central to interpreting the polarisation signal. When
two poles contribute comparable fluxes but different
linear PDs, their combination dilutes the more strongly
polarised component, even if both are dominated by the
same mode. Stronger cancellation occurs when
contributions dominated by different modes are combined,
either between the two poles or across energies within
a band. Accordingly, in Fig.~\ref{fig:polaris}, the
broad 2--8\,keV band has an overall lower PD than
either sub-band.

As we assume symmetric antipodal poles of a dipolar
field, their position angles differ by $180^\circ$,
$\chi_2=\chi_1+180^\circ$, and therefore follow the
behavior of the classical RVM. In the
absence of a global dominant mode change during the rotation, the
PA of mode~2 follows the RVM for
emission from either pole, as illustrated by the
PA panels for the $2$--$4\kev$ band.
Among the observables considered here, only the
PA distinguishes between the two
configurations corresponding to a given pair
$(i_1,i_2)$: $\Theta_1=i_1$ and $i=i_2$, or
$\Theta_1=i_2$ and $i=i_1$. We therefore show the
PA for both configurations in
Fig.~\ref{fig:polaris}. When mode~1 becomes dominant,
the PA switches to the orthogonal RVM
branch, producing a $90^\circ$ jump. One mode may
dominate throughout the entire cycle, as for
geometry~II in the 4--8\,keV band, or repeated
changes in mode dominance may produce transitions
between the classical and orthogonal RVM branches,
resulting in a complex polarisation-angle behavior
(e.g., III and IV).

The apparent discontinuities in the PA
curves therefore have two distinct origins. Some are
ordinary consequences of wrapping the angle modulo
$180^\circ$, for example from $-90^\circ$ to
$+90^\circ$. Others are physical $90^\circ$ jumps caused
by a change in the globally dominant mode within the
energy band. The latter are accompanied by a decrease
in the PD to nearly zero, where the
PA becomes poorly defined. If such a
transition occurs rapidly, as for geometry~IV in the
$4$--$8\kev$ band, averaging over finite phase bins may
also produce apparent PA changes that
differ from $90^\circ$.

Finally, general relativistic effects remain important
for modelling the polarisation signal even for the
slowly rotating X-ray pulsars considered here. Although
the polarisation plane does not rotate and the PD is
conserved along each geodesic
\citep[e.g.,][]{poutanen2020}, light bending affects
the observed polarisation in two ways. First, at a
given rotational phase, it allows the observer to sample
smaller emission angles to the magnetic field than in
flat spacetime. Because the modes become increasingly
circular towards the field direction, this reduces the
observed linear PD. This effect is expected to strengthen
with increasing neutron star compactness. Second, when
several emitting regions are visible, light bending
changes their relative contributions to the observed
Stokes parameters through different lensing factors.
These factors do not cancel when the contributions are
combined and therefore affect both the PA and PD (e.g., Eq.~\ref{eq:pdmix}).

\section{Summary and Outlook}
\label{sec:ende}

We constructed a steady-state emission model for thermal
Comptonisation in dense, strongly magnetised plasma at the
base of an X-ray pulsar accretion channel, which
captures the principal beaming, polarisation, and
energy-redistribution effects in the continuum and
cyclotron line.
Combined
with relativistic, phase-dependent projection of the
emission onto the observer's sky, the model predicts
observables without specifying a particular luminosity state.

We found that thermal Comptonization leaves distinct
fingerprints in the continuum, cyclotron lines,
polarisation, and their pulse-phase variability. Many
arise from the interplay between the magnetised plasma
and QED vacuum, which together determine photon propagation.
We investigated the principal observable signatures
across a broad range of physical parameters
typical of the base of an accretion channel
and explained their
evolution with neutron star geometry. This evolution is
driven primarily by the range of emission angles sampled
over the rotational cycle, together with the angular
dependence of the emission itself. While some observables
are highly sensitive to the physical conditions, others
remain robust and can therefore provide a reliable
characterisation of the neutron star geometry and the
dominant emission process. The most important results are:

 \begin{itemize}

    \item[--] The emitted beams resemble standard
    ``pencil'' (slab/spot) and ``fan'' (cylinder/column) patterns
    only at intermediate X-rays well below $E_\cyc$. Their
    shapes vary with magnetic field,
    but remain featureless across the studied range of
    magnetic fields. Soft-X-ray beams from a spot often show a
    split of ${\sim}20^\circ$ around the field direction,
    which widens with
    increasing vacuum influence and lower $\tau_\T$. At $E_\cyc$,
    both region shapes produce narrow central beams and side
    petals. In the line wings, the central beam and petals
    strengthen for the column and spot, respectively. Their
    relative strength and width
    reflect the interplay between plasma and vacuum effects,
    while the overall morphology remains robust across the
    explored physical parameters.

    \item[--] Intermediate-energy pulse profiles for hot spots
    primarily trace the geometry and number of
    visible poles, retaining a similar morphology across
    the studied $E_\cyc$ values. When
    the soft-X-ray beam split, additional peaks arise when a pole
    is viewed within $\lesssim20^\circ$ of the magnetic-field
    direction, making the number of peaks in these profiles
    potentially an unreliable tracer of number of emitting
    regions. In the cyclotron line band, the
    central beam and side petals can produce interpulses
    and inversions in profiles, varying with $E_\cyc$.

    \item[--] Column pulse profiles shows
    minima at the closest approach of the pole; the
    two separate peaks arise only for high-angle
    geometries, where each column is occulted during
    part of the cycle. Within each energy band, their
    morphology remains similar across
    studied $E_\cyc$, except at the lowest one.

    \item[--] The interplay between the central beam and
    side petals near the cyclotron line produces diverse
    structures in the pulsed fraction spectra of hot-spot
    emission, including bumps, dips, and W- and M-shaped
    modulations.
    For column emission, geometry-dependent visibility
    produces dips in small-angle configurations and
    pronounced bumps otherwise.

    \item[--] The resulting cyclotron-line widths are
    unreliable plasma-temperature diagnostics. In
    phase-averaged spectra, they depend on geometry
    and usually exceed standard Doppler-width predictions,
    leading to much higher $kT_\el$ inferred in
    this way. For column emission, however, they can
    fall below this prediction. This
    extreme narrowing disappears towards lower $E_\cyc$.
    Together, this further disfavours a (thermal)
    electron-cyclotron origin for narrow features in soft
    X-rays found in ultraluminous X-ray sources
    \citep[e.g.,][]{brightman2018}, but remains to
    be verified at lower fields.

\item[--] The soft-X-ray PD is generally $\lesssim30\%$
    below $E_\cyc$ and decreases further through dilution
    and cancellation when integrated over wider energy
    ranges, helping to explain the low PDs measured by
    \textit{IXPE} \citep{poutanen2024}. The plasma
    contribution drives the polarisation modes away
    from linearity already at large angles to the magnetic
    field, while their circular component increases
    towards the field, which can substantially reduce
    the observed linear PD. Its omission may partly
    explain the high linear PD predicted by
    \citet{caiazzo2020a}, whose treatment of internal
    column emission shows circular
    polarisation mainly at $E_\cyc$ or
    very close to the field direction. Their single-mode
    emission assumption may also contribute; our model
    predict that column also emits in two modes. By
    allowing smaller angles to the field to be sampled,
    light bending may further enhance this plasma-induced
    depolarisation as compactness increases.

\item[--] Observable PA behaviour depends on whether
    mode changes survive energy integration. Localised
    changes caused by mode crossings or the vacuum
    resonance can be largely washed out, whereas
    pronounced PD minima and $90^\circ$ PA jumps arise
    when a single-pole beam or changing pole
    contributions allow the other mode to dominate
    over a broad phase interval. Otherwise, the PA
    follows the RVM or its orthogonal branch. Geometry,
    energy, and physical parameters can thus produce
    RVM-like or more complex behaviour, qualitatively
    matching the diversity observed by \textit{IXPE}
    across sources and energy bands
    \citep[e.g.,][]{forsblom2025,zhao2026}. Phase-dependent
    PA may thus vary between observations with the physical
    state of the emission region rather than geometry,
    making it a less robust geometry diagnostic, unless
    magnetospheric vacuum polarisation, omitted here
    \citep{heyl2000}, largely suppresses this sensitivity.

 \end{itemize}

Several effects that may influence the observables
are omitted here: for example, relativistic corrections
to the cyclotron resonances, multi-photon processes,
and emission-region inhomogeneity. A fully relativistic
treatment is expected to produce a somewhat narrower,
more asymmetric cyclotron line \citep{alexander1989,bussard1986},
together with small angle-dependent shifts in the
resonance energy. The importance of multi-photon
processes depends on the interaction probabilities
set by $E_\cyc$ and the continuum flux near the
resonances. They may be non-negligible in our 
low-field models, particularly for $E_\cyc=20\kev$;
this case should therefore be treated cautiously,
only as a reference within our model framework.

Unlike continuum modeling, detailed Monte Carlo
simulations of cyclotron lines typically include
multi-photon processes and rely on full QED
treatment, while traditionally employ linear photon
normal modes determined by QED vacuum polarisation
(e.g., \citealt{araya1999, schonherr2007, schwarm2017a,
schwarm2017b, kumar2022} based on \citealt{shabad1975, sina1996}).
Vacuum polarisation has therefore long been included
in determining the photon modes in this class of
models, whereas the magnetised plasma contribution
has generally been neglected.
Our results demonstrate the need to include this
contribution, particularly at relatively low magnetic
fields, where several cyclotron harmonics fall within
the X-ray band, and revise previous results of
detailed cyclotron-line simulations.

The effects of inhomogeneity are less straightforward 
to assess. The constant density is likely the most
limiting assumption, given the strong density gradient
expected even across the small vertical extent of
the region in the strong gravitational field.
Lower-density upper
layers may preserve the soft-X-ray beam split over a
wider range of physical parameters. They may also
alter the location and extent of the vacuum-resonance
region, as well as a more realistic treatment of partial
mode conversion in its vicinity
\citep[e.g.,][]{van_adelsberg2006}. Our results suggest
that band or phase averaging may largely wash out these
effects, although lower-density outer layers could
increase linear polarisation at intermediate angles
to the magnetic field.

In a forthcoming study (Sokolova-Lapa et al., in prep.),
we map these observables across geometries, combine them
into joint diagnostics, and investigate their
morphological and statistical properties. We then apply
these diagnostics to observations to assess the model’s
applicability to real sources and constrain neutron star
geometry, types of emission regions, and the
spectral-formation mechanism.

\begin{acknowledgements}
  This research was supported by the International
  Space Science Institute (ISSI) in Bern, through
  ISSI International Team projects \#495, ``Feeding
  the spinning top'' and 25-657, ``Polarimetric
  Insights into Extreme Magnetism'' and through
  the Working Group project ``Disentangling Pulse
  Profiles of (Accreting) Neutron Stars''. ESL
  acknowledges support from the FAU Emerging Talents
  Initiative and funding from the European Union’s
  Horizon 2020 programme under the AHEAD2020 project
  (grant agreement No. 871158). GL acknowledges 
  support by the Deutsche Forschungsgemeinschaft
  (DFG) under project number 570950648.
\end{acknowledgements}

\clearpage
\begin{appendix}

\section{Model availability}
\label{sec:data}
The \texttt{sigel-T} local-emission tables are available
at \url{https://www.sternwarte.uni-erlangen.de/research/sigel-t}
(“T” stands for thermal Comptonization). They use the OGIP
\texttt{XSPEC} table-model FITS structure for convenient
storage and interpolation. The tabulated quantities are
angle-dependent, bin-integrated photon specific intensities
in the local emission-region frame, not observer-frame fluxes.

\section{Polarisation}
\label{sec:appa}

\subsection{Computational approach}

We calculated all polarisation quantities first as
functions of energy and rotational phase and subsequently
performed the required integrations. For a given geometry,
we calculated the angle $\psi_k$ between the LOS and the
magnetic axis of pole $k$ at each rotational phase,
\begin{equation}
    \cos\psi_k=\cos i\cos\Theta_k+\sin i\sin\Theta_k\cos(\phi+\phi_k),
\end{equation}
where $\phi_k$ is the azimuthal phase offset
($\phi_1=0$ and $\phi_2=\pi$ for the antipodal poles
considered here). This determines the corresponding local
emission angle, $\theta_k$, of radiation reaching the
observer from each visible pole, as in Figs.~\ref{fig:ppem_spot}
and \ref{fig:ppem_cyl}. Here, however, we adopted the
analytical light-bending approximation of
\citet{poutanen2020a}, which is sufficiently accurate
for surface emission and computationally efficient.
The corresponding lensing factor is
\begin{equation}
    \mathcal D_k=\frac{1}{1-\mathcal C}\frac{{\diff}\cos\theta_k}{{\diff}\cos\psi_k},
\end{equation}

At a given observed energy $E$, we interpolated the
internal specific intensities of each polarisation mode
$j$ at $\theta_k$ and the emitted energy $(1+z)E$,
obtaining $I_j^k[\theta_k,(1+z)E]$. Together with the
signed linear-polarisation fractions $p_{q,j}$, these
intensities give $Q^k_\mathrm{loc}$ through
Eq.~\ref{eq:qstock}. In the local
basis, $U^k_\mathrm{loc}=0$. We did not calculate
the Stokes V as in this work we focused on linear
polarisation.

Projection, light bending, and gravitational redshift
result in the flux weight
\begin{equation}
    \mathcal W_k(\phi)=\frac{\cos\theta_k\,\mathcal D_k}{(1+z)^3},
\end{equation}
which must be applied to $I$ and $Q_\mathrm{loc}$.
Similar to \citet{meszaros1988a}, we determined the
projection of the magnetic axis onto the observer's
sky, $\chi_k$, using RVM and rotate the polarisation
basis. The observer-frame Stokes
parameters for each pole are
\begin{equation}
    \begin{aligned}
        I^k_\mathrm{obs}
            &=\mathcal W_k I^k_\mathrm{loc},\\
        Q^k_\mathrm{obs}
            &=\mathcal W_k Q^k_\mathrm{loc}\cos(2\chi_k),\\
        U^k_\mathrm{obs}
            &=\mathcal W_k Q^k_\mathrm{loc}\sin(2\chi_k).
    \end{aligned}
\end{equation}
The orthogonal orientations of the modes are incorporated
through the opposite signs of $p_{q,1}$ and $p_{q,2}$,
which is equivalent to assigning mode $chi$ differing by
$90^\circ$. Contributions from the visible poles were
then added in the observer frame,
\begin{equation}
    (I_\mathrm{tot},Q_\mathrm{tot},U_\mathrm{tot})=\sum_k
    (I^k_\mathrm{obs},Q^k_\mathrm{obs},U^k_\mathrm{obs}).
\end{equation}

For phase- and energy-resolved quantities, the total
PD can therefore be written explicitly as
\begin{equation}
\label{eq:pdmix}
    \mathrm{PD}=\frac{\sqrt{
    \left[ \sum_{k=1}^{2} \mathcal W_k Q^k_\mathrm{loc}\cos(2\chi_k)\right]^2 + 
    \left[\sum_{k=1}^{2}\mathcal W_k Q^k_\mathrm{loc}\sin(2\chi_k)\right]^2}}
    {\sum_{k=1}^{2}\mathcal W_k I^k_\mathrm{loc}}.
\end{equation}
For finite energy bands and, where applicable, phase
intervals, we first integrated the Stokes parameters
and then calculated the polarisation observables. We
retained the same notation for the resulting integrated
quantities. The PA and PD are
\begin{equation}
    \mathrm{PA}=\frac{1}{2}\operatorname{atan2} \left( U_\mathrm{tot},Q_\mathrm{tot} \right)
\end{equation}
and
\begin{equation}
    \mathrm{PD}=
    \frac{\sqrt{Q_{\mathrm{tot}}^2+U_{\mathrm{tot}}^2}}
         {I_{\mathrm{tot}}}.
\end{equation}

To trace mode dominance independently of the intrinsic
linear polarisation of the modes, we defined the signed
mode difference for a pole $k$
\begin{equation}
\label{eq:ck}
    C_k= \frac{ I_{\mathrm{obs},1}^k-I_{\mathrm{obs},2}^k} {I_\mathrm{tot}}.
\end{equation}
The total signed mode difference is therefore
\begin{equation}
\label{eq:ctot}
    C_\mathrm{tot}=C_1+C_2 = \frac{\displaystyle\sum_{k=1}^{2} \left(I_{\mathrm{obs},1}^k-I_{\mathrm{obs},2}^k\right)}{I_\mathrm{tot}}.
\end{equation}

\subsection{Phase-averaged polarisation degree}

\begin{figure}
  \centering
  \includegraphics[width=\columnwidth]{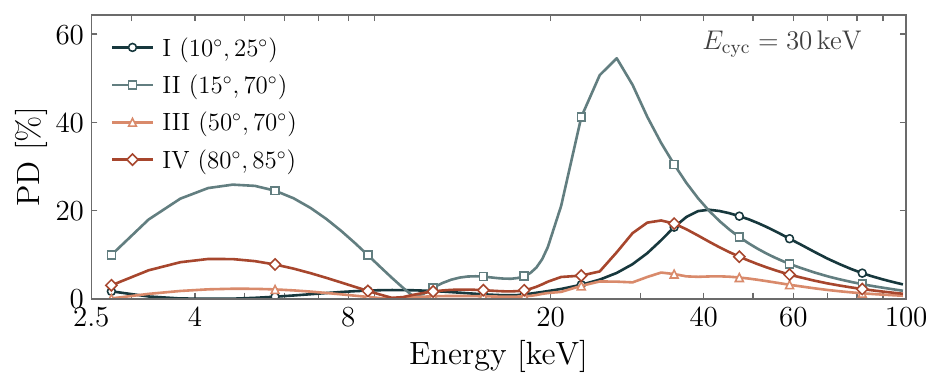}
  \includegraphics[width=\columnwidth]{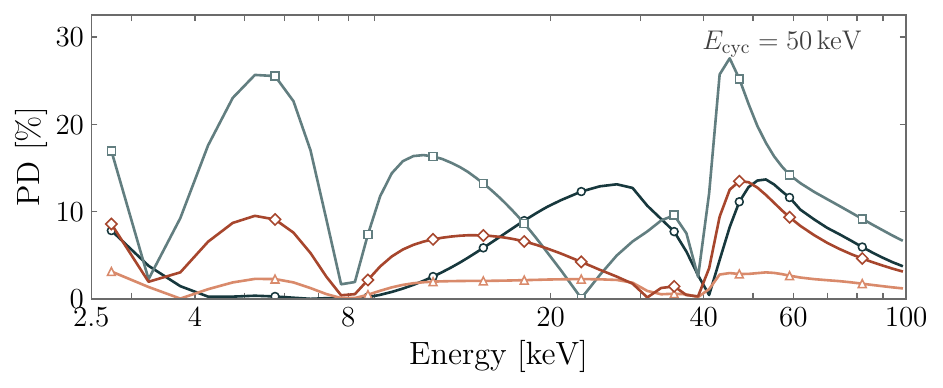}
  \includegraphics[width=\columnwidth]{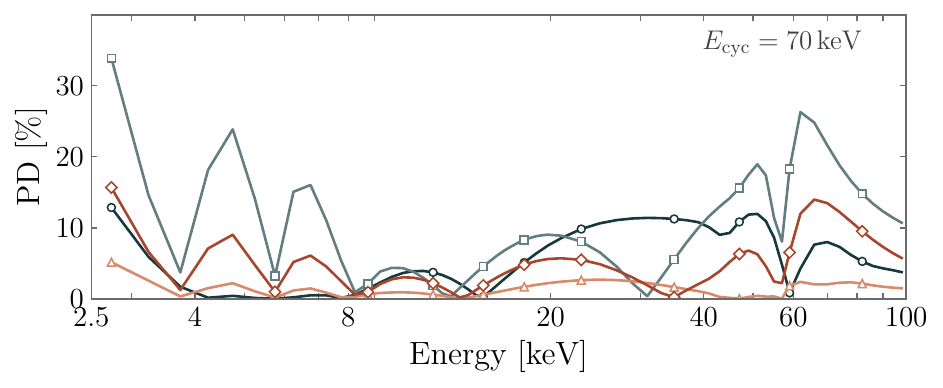}
  \caption{Phase-averaged PD spectra for geometries I--IV
  and $E_\cyc=30,\mathrm{keV}$ (top), the reference model
  with $E_\cyc=50,\mathrm{keV}$ (middle), and
  $E_\cyc=70,\mathrm{keV}$ (bottom).}
\label{fig:pden}
\end{figure}

Besides the energy-averaged, phase-resolved quantities
shown in Fig.~\ref{fig:polaris}, we computed the
phase-integrated $\mathrm{PD}(E)$. These curves provide
a convenient overview of the PD range produced by
different geometries and magnetic-field strengths.
Figure~\ref{fig:pden} shows $\mathrm{PD}(E)$ for three
internal-emission models with different magnetic
fields. Away from the cyclotron line, the PD
generally remains $\lesssim20$--$30\%$. For the reference
model (middle panel), the curves broadly follow the
energy-dependent signed mode difference of the internal
emission shown in the bottom panel of
Fig.~\ref{fig:emprof}. Minima in $\mathrm{PD}(E)$ may
therefore retain signatures of continuum mode crossings
or the vacuum resonance, depending on geometry.
  
\section{Doppler broadening of the cyclotron resonance}
\label{sec:appc}

\subsection{Internal emission}
\begin{figure}
    \includegraphics[width=\columnwidth]{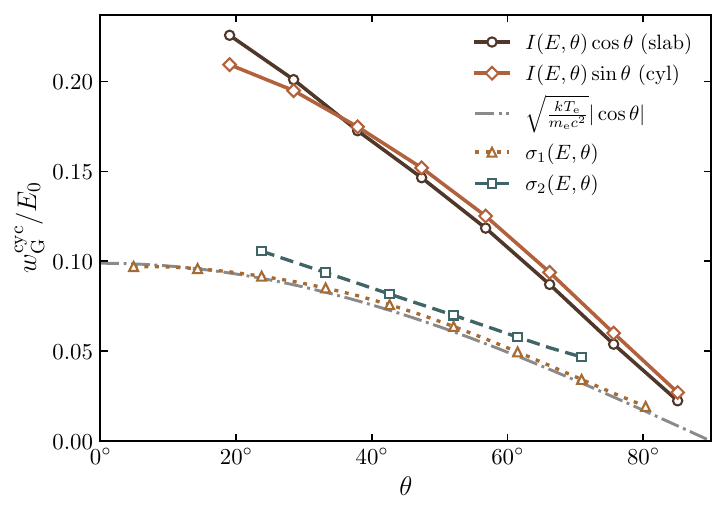}
    \caption{Dependence of the cyclotron-line broadening,
    quantified by the Gaussian width-to-centroid
    ratio $w^\cyc_\mathrm{G}/E_0$, on the emission angle $\theta$
    for our reference models (see Fig.~\ref{fig:emprof}) for
    the slab and cylinder (open circles and diamonds, respectively).
    See details in the text (Sect.~\ref{sec:appc}) The grey dash-dotted
    line gives the analytical prediction for one-dimensional thermal
    Doppler broadening. Open squares (mode~1) and triangles (mode~2)
    show the corresponding ratios obtained by fitting a power law
    plus a Gaussian emission profile to the scattering cross sections.
  }
\label{fig:sigma}
\end{figure}

To quantify the width of the cyclotron line and
compare its angular dependence with the classical
Doppler width for one-dimensional Maxwellian electrons
(e.g., MN85a),
\begin{equation}\label{eq:fwhm}
    \frac{w_\mathrm{G}(\theta)}{E_\cyc} = \sqrt{\frac{kT_\el}{m_\el c^2}}|\cos\theta|,
\end{equation}
we fitted he angle-dependent differential-flux spectra,
summed over the two polarisation modes (see Fig.~\ref{fig:emprof}
between $10$ and $100\kev$ with a cutoff power law multiplied
by a Gaussian absorption profile,
\begin{equation}
    F(E) = A\,E^{-\Gamma} \exp\!\left(-\frac{E}{E_{\mathrm{fold}}}\right)
    \exp\!\left[ -\tau_\mathrm{G} \exp\!\left( -\frac{(E-E_0)^2}{2{w^\cyc_\mathrm{G}}^2} \right) \right].
\end{equation}
Here $A$, $\Gamma$, and $E_\mathrm{fold}$ describe the
continuum, while $E_0$, $w^\cyc_\mathrm{G}$, and
$\tau_\mathrm{G}$ are the line centroid, Gaussian standard
deviation, and line-centre optical depth, respectively.
We have also verified the general validity of obtained
fits by visual inspection. We omitted the fits at
$\theta\simeq0^\circ$ for both media and at
$\theta\simeq90^\circ$ for the slab, where the emergent
spectra contain no identifiable cyclotron absorption
line and the Gaussian component instead follows broader
spectral curvature. We separately characterised broadening
of the resonances in the cross sections of both polarisation
modes by fitting the power law continuum with the
additive Gaussian emission profile around the
cyclotron energy. We present values only over
angular ranges where the cyclotron resonance
in the cross sections can be adequatly described
by the single Gaussian (i.e., are not strongly
asymmetric). Figure~\ref{fig:sigma} shows the
resulting broadening ratios, $w^\cyc_\mathrm{G}/E_0$,
which we adopt as descriptive measure of the line.
See the related discussion in Sect.~\ref{sec:res:sigelt:cyc}.

\subsection{Phase-averaged spectra}
\begin{figure*}
  \centering
  \includegraphics[width=0.49\textwidth]{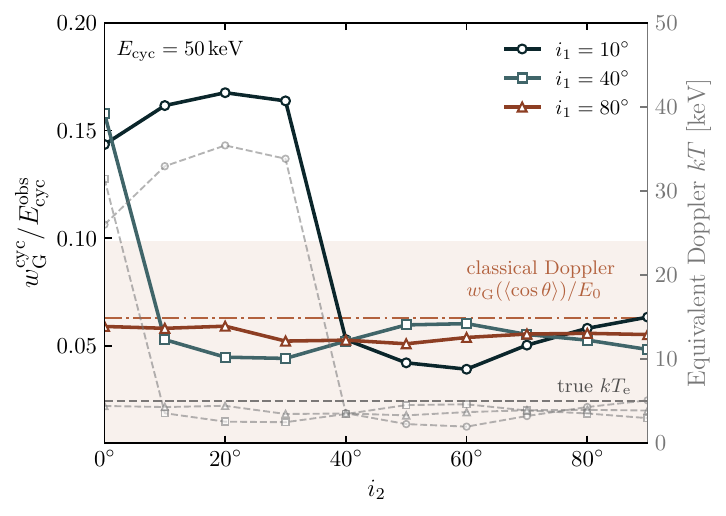}
  \hfill
  \includegraphics[width=0.49\textwidth]{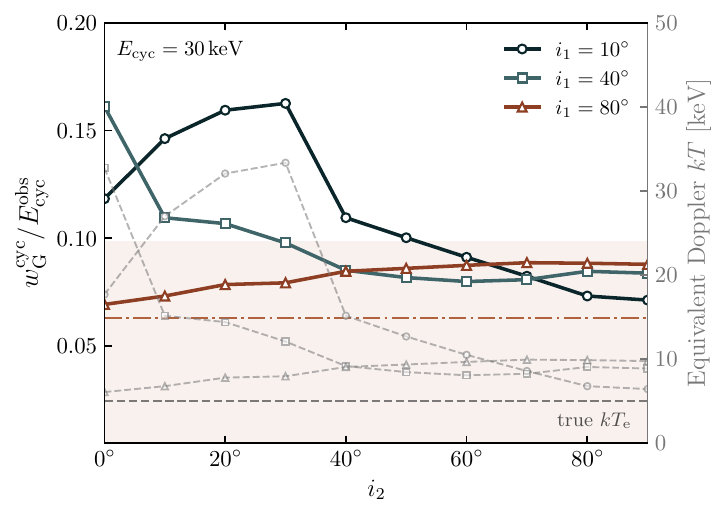}
  \caption{Same as in Fig.~\ref{fig:sigma_phav}, but for column emission, $h=300\,\mathrm{m}$, for the reference model (left) and its variance with $E_\cyc=30\,\mathrm{keV}$ (right).
  }
\label{fig:sigma_phav_cyl}
\end{figure*}

We applied the same fitting procedure to the
phase-averaged spectra presented in
Sect.~\ref{sec:res:lbsigelt:flux}. The results for
the reference model and the hot spot emission is
presented in Fig.~\ref{fig:sigma_phav}.
Figure~\ref{fig:sigma_phav_cyl} also shows the
width behaviour for the column case. The results
are discussed in Sect.~\ref{sec:res:lbsigelt:flux}.

\section{Effect of photon normal modes choice on beaming}
\label{sec:appd}
Due to the pronounced differences between approximations
for photon normal modes adopted in various models, it is
important to determine their principal effects. For the
observables considered here, the shapes of the emitted
beams at different energies, together with the
geometry-specific visibility conditions, determine much
of their behaviour. Recently, \citet{markozov2026b}
showed that plasma-defined normal modes produce
substantially different beam profiles compared to when vacuum polarisation is also included,
together with plasma. Many existing models, particularly
detailed cyclotron-line simulations, nevertheless adopt
either pure vacuum
normal modes (see Sect.~\ref{sec:ende}) or simply strictly
transverse linear modes. The latter approximate
the vacuum modes at magnetic fields below the Schwinger limit,
$B\lesssim B_\mathrm{c}\simeq4.4\times10^{13},\mathrm{G}$.

\begin{figure*}
    \centering
    \includegraphics[width=.32\textwidth]{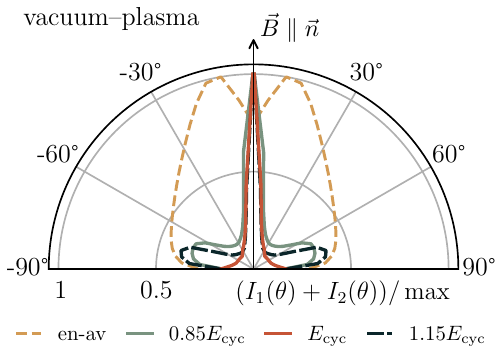}
    \includegraphics[width=.32\textwidth]{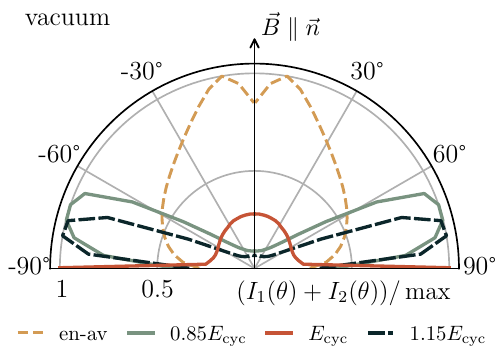}
    \includegraphics[width=.32\textwidth]{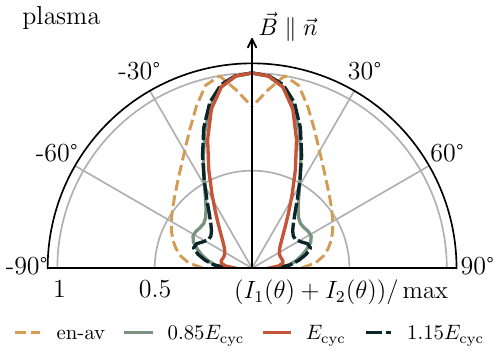}
    \caption{Beam patterns in the vicinity of the cyclotron resonance for     
    different types of photon polarisation modes: mixed plasma--vacuum contribution, as used throughout this work (left), pure vacuum (middle), pure plasma (right). The patterns are given for the
    model $E_\cyc=50\kev$, $kT_\el=5\kev$, $\tau_\T=50$, and $n_\el=10^{23}\,\mathrm{cm}^{-3}$.}
    \label{fig:vacpl}
\end{figure*}

\begin{figure*}
    \centering
    \includegraphics[width=0.95\textwidth]{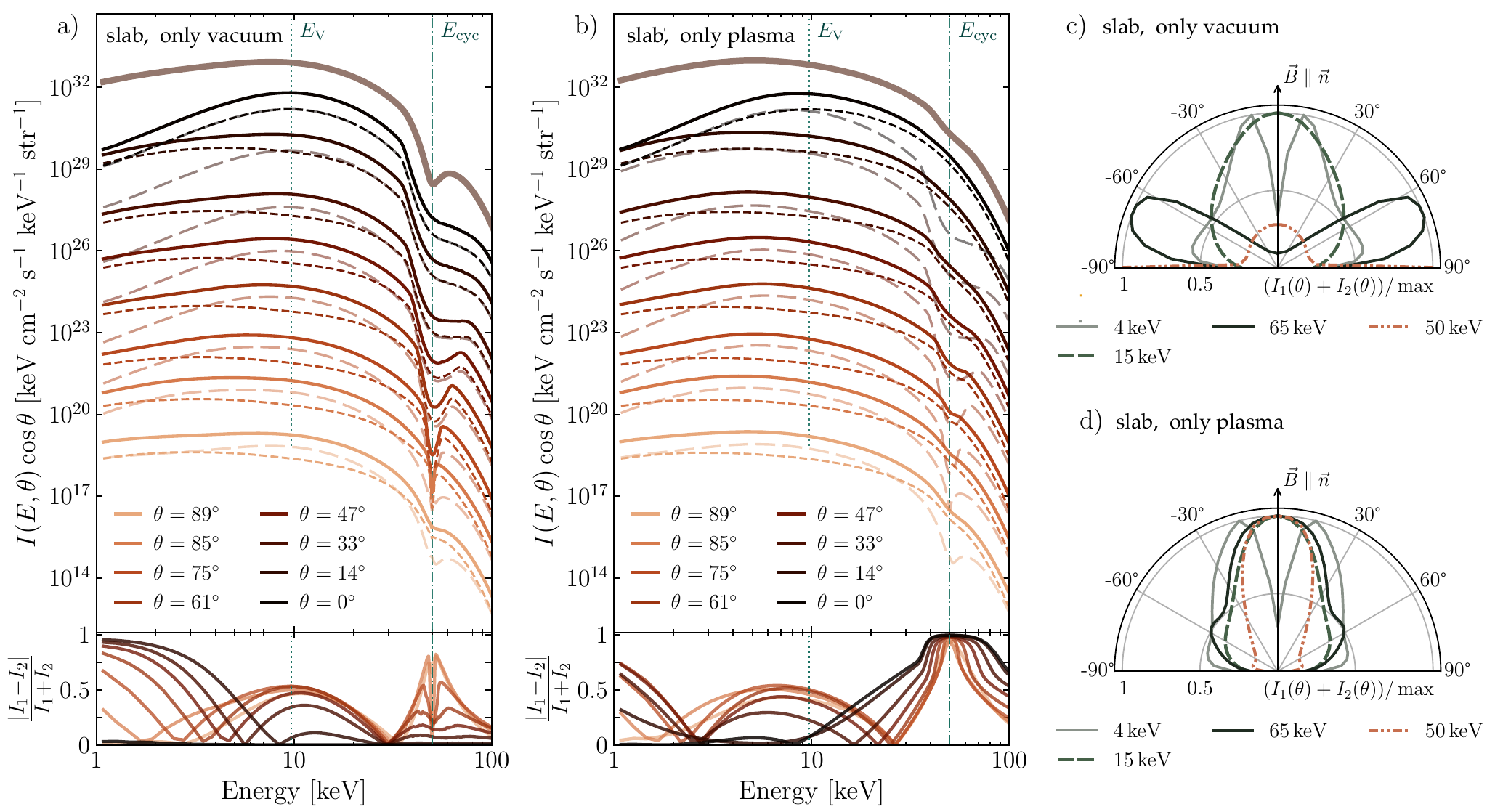}
    \caption{Same as in Fig.~\ref{fig:emprof}, but only
    for the slab-like emission region of $\tau=50$, with
    pure vacuum (a,c) and pure plasma (b, d) photon
    polarisation modes.}
    \label{fig:emprof_plvac}
\end{figure*}

To isolate these effects, we compare the mixed
plasma--vacuum modes used here with the pure-vacuum
and pure-plasma cases. Beam patterns
near the cyclotron energy, as well as those integrated
over the full energy range, are shown in
Fig.~\ref{fig:vacpl}. While the broad-band profiles are
only weakly affected, the beaming near the cyclotron
resonance changes dramatically. The mixed profiles
appear to resemble a combination of the two pure cases,
with the side petals arising from the vacuum contribution
and the central beam defined by the plasma. Figure~\ref{fig:emprof_plvac}
additionally shows
angle-dependent spectra for both pure cases and emission
profiles at the same energies as in Fig.~\ref{fig:emprof}. See the
related discussion in Sect.~\ref{sec:res:sigelt:beam}.

\section{Examples of beam profiles and spectra for different physical parameters}
\label{sec:appe}

To extend our results presented in Secs.~\ref{sec:res:sigelt:specf}
and \ref{sec:res:sigelt:beam}, here we show two more models with
different magnetic fields.

\begin{figure*}
\centering
\begin{subfigure}{\textwidth}
    \centering
    \includegraphics[width=\textwidth]{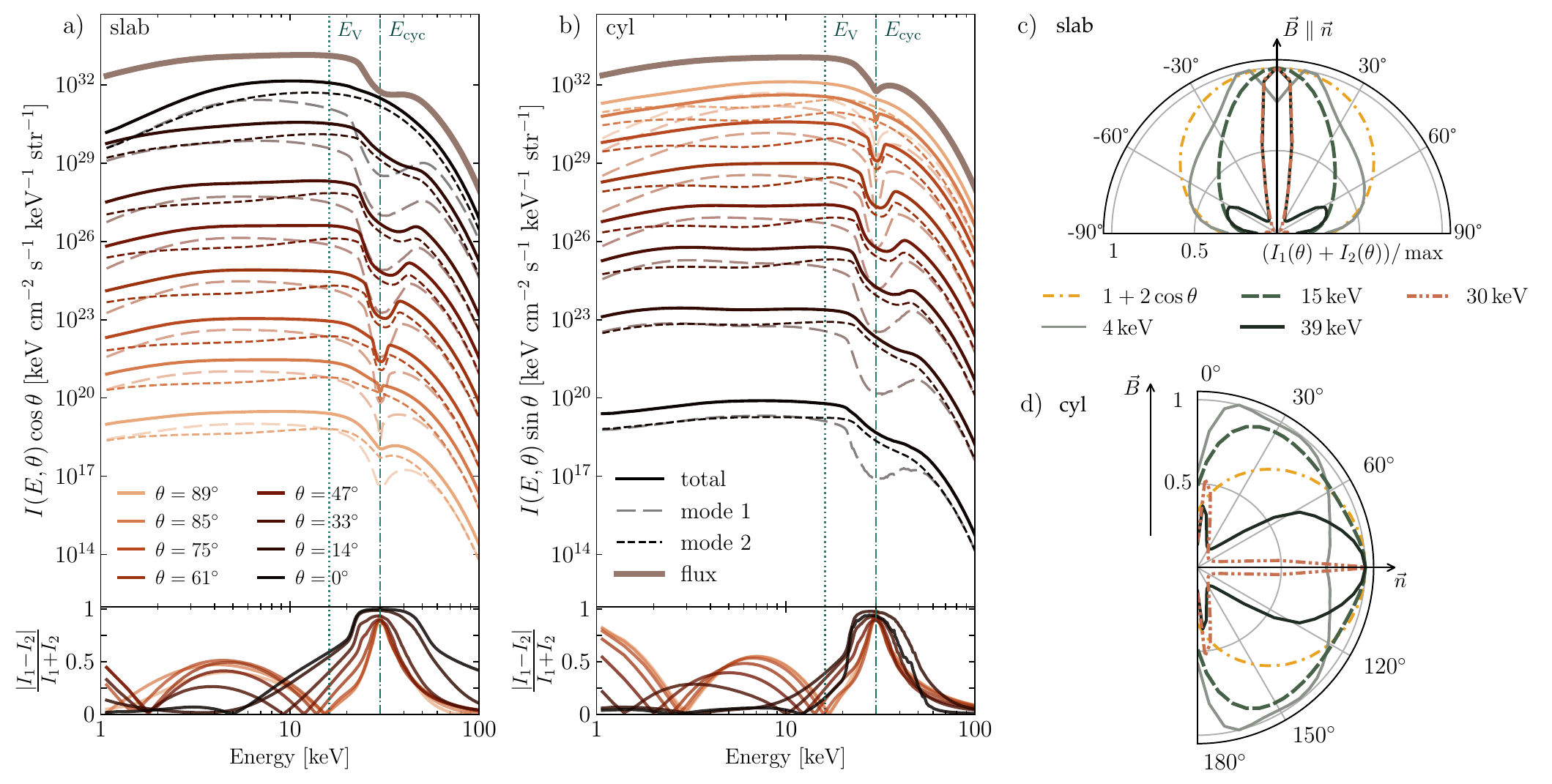}
    \caption{$E_\cyc=30\,\mathrm{keV}$, $kT_\el=5\,\mathrm{keV}$, $n_\el=10^{23}\,\mathrm{cm}^{-3}$, and $\tau_\T=50$ for both shapes of the emission region.}
    \label{fig:beam_a}
\end{subfigure}

\vspace{0.4cm}

\begin{subfigure}{\textwidth}
    \centering
    \includegraphics[width=\textwidth]{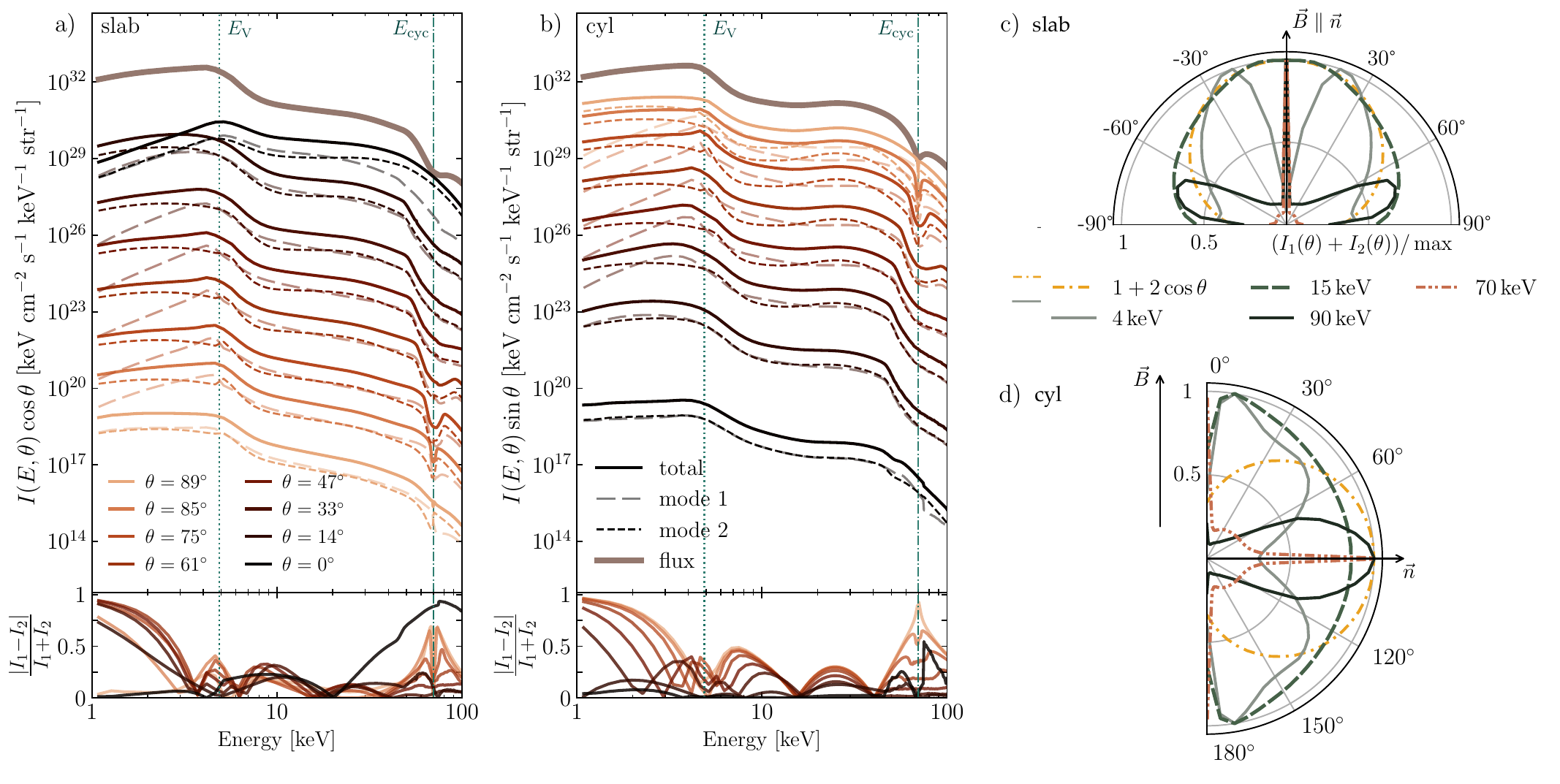}
    \caption{$E_\cyc=70\,\mathrm{keV}$, $kT_\el=8\,\mathrm{keV}$, $n_\el=5\times10^{22}\,\mathrm{cm}^{-3}$, and $tau_\T=20$ for the slab and $\tau_\T=50$ for the cylinder.}
    \label{fig:beam_b}
\end{subfigure}

\caption{Same as in Fig.~\ref{fig:emprof} but for different physical parameters of the emitting region}
\label{fig:beam_app}
\end{figure*}

\end{appendix}

\end{document}